\documentclass{jfm}
\usepackage{graphicx}
\usepackage{newtxtext}
\usepackage{newtxmath}
\usepackage{natbib}
\usepackage{hyperref}
\usepackage{bm}
\usepackage{caption}
\usepackage[dvipsnames]{xcolor}
\hypersetup{
    colorlinks = true,
    urlcolor   = blue,
    citecolor  = CadetBlue,
}

\title[Transient and steady convection in two dimensions]{Transient and steady convection in two dimensions}

\author{Ambrish Pandey\aff{1,2}
        \corresp{\email{ambrish.pandey@ph.iitr.ac.in}}
        \and Katepalli R. Sreenivasan\aff{2,3}
        }

\affiliation{\aff{1} Department of Physics, Indian Institute of Technology Roorkee, Roorkee 247667, Uttarakhand, India
\aff{2} Center for Astrophysics and Space Science, New York University Abu Dhabi, Abu Dhabi 129188, United Arab Emirates \\
\aff{3} Tandon School of Engineering, Department of Physics, and Courant Institute of Mathematical Sciences, New York University, New York, NY 11201, USA
}

\begin{document}
\maketitle

\begin{abstract}
We simulate thermal convection in a two-dimensional square box using the no-slip condition on all boundaries, and isothermal bottom and top walls and adiabatic sidewalls. We choose 0.1 and 1 for the Prandtl number $Pr$ and vary the Rayleigh number $Ra$ between $10^6$ and $10^{12}$. We particularly study the temporal evolution of integral transport quantities towards their steady states. Perhaps not surprisingly, the velocity field evolves more slowly than the thermal field, and its steady state---which is nominal in the sense that large-amplitude low-frequency oscillations persist around plausible averages---is reached exponentially. We study these oscillation characteristics. The transient time for the velocity field to achieve its nominal steady state increases almost linearly with the Reynolds number. For large $Ra$, the Reynolds number itself scales almost as $Ra^{2/3} Pr^{-1}$, and the Nusselt number as $Ra^{2/7}$.
\end{abstract}


\section{Introduction} 
\label{sec:intro}

Turbulent flows driven by buoyancy due to inhomogeneity of the temperature are common in nature and applications~\citep{Verma:book2018, Schumacher:RMP2020, Lohse:PT2023}. Rayleigh-B\'enard convection (RBC) is a paradigm for such flows. The RBC originally referred to shallow horizontally extended layers of fluid, heated from below and cooled from above, and the horizontal walls are smooth unless otherwise specified. In this traditional paradigm, RBC is entirely governed by Prandtl and Rayleigh numbers---where the Prandtl number $Pr$ is the ratio of the kinematic viscosity $\nu$ to the thermal diffusivity $\kappa$ of the fluid, and the Rayleigh number $Ra$ is the ratio of the forcing strength to the dissipative mechanisms.  Heat and momentum transport across the convective fluid layer are two global responses to thermal driving in RBC. Heat transport is measured by the Nusselt number $Nu$, which is the total heat flux relative to that by conduction in the absence of fluid motion, and the momentum transport by an appropriate Reynolds number $Re$, which defines the flow strength. 

Because the Rayleigh number is proportional to $H^3$, where $H$ is the height of the convection apparatus, there has been a tendency in the last 25 or so years to choose as high a value of $H$ as possible while, by necessity, shrinking the horizontal dimension (e.g., \citet{Castaing:JFM1989, Niemela:Nature2000}). The same is also true of direct numerical simulations (DNS)~(e.g., \citet{Stevens:JFM2011, Iyer:PNAS2020}). The choice of a low aspect ratio $\Gamma \equiv L/H$ (where $L$ is the horizontal dimension of the apparatus) is common in the quest to achieve very high Rayleigh numbers, but an organized motion that develops in such flows has its own structural morphology~\citep{Kadanoff:PT2001, Sreenivasan:PRE2002,Chilla:EPJE2012, Foroozani:PRE2014, Foroozani:PRE2017, Verma:NJP2017} that depends on the aspect ratio and the shape of the apparatus; see also \citet{Pandey:PD2022, Stevens:JFM2024}. Pure scaling laws in such flows are unlikely for all conditions, yet it is common to fit the Nusselt and Reynolds numbers by power laws with respect to $Ra$, i.e., $Nu \sim Ra^{\gamma}$ and $Re \sim Ra^{\zeta}$. As pointed out by \citet{Doering:PRL2020}, fitting such local exponents for small ranges of data is bound to lead to conclusions of uncertain value. Indeed, there are considerable variations of the effective exponents $\gamma$ and $\zeta$  from one study to another, and have been the subject of extensive reviews---e.g., by \citet{Chilla:EPJE2012, Lohse:RMP2024}.  In a series of large simulations, an effort has been made to avoid the constraining effect of the sidewalls by stipulating periodic boundary conditions on them~\citep{Samuel:JFM2024}. These studies mimicking large aspect ratio convection have revealed a very different nature of near-wall velocity from a traditional boundary layer that undergoes laminar-turbulent transition, and have implications for the so-called ultimate state.

There has been the expectation that the riddle of the ultimate state can be solved for the two-dimensional (2D) case in which the flow is compelled to occur only in a vertical plane~\citep{Zhu:PRL2018, Samuel:PRF2024, Tiwari:IJHMT2025} because of the natural hope that very high Rayleigh numbers can be achieved here for the same computing power. Existing data show a heat transport scaling that is similar to the three-dimensional (3D) counterpart~\citep{Schmalzl:EPL2004, Poel:JFM2013, Pandey:Pramana2016, Zhang:JFM2017, Pandey:JFM2021} but the Nusselt number in 2D is smaller when $Pr \geq 1$~\citep{Poel:JFM2013}, although essentially the same when $Pr$ is small~\citep{Pandey:JFM2021}, as for liquid metals. The Reynolds number, on the other hand, shows very different behaviors in 2D and 3D. The magnitudes of momentum transport and the scaling exponent $\zeta$ are consistently higher in 2D, with $\zeta \geq 0.60$~\citep{Schmalzl:EPL2004, Poel:JFM2013, Zhang:JFM2017, Wan:JFM2020, Pandey:JFM2021}. As a summary, the exponent $\gamma$ has been reported to take values in the range [1/4, 1/3], while $\zeta$ assumes values in the range [4/9, 2/3]~\citep{Verma:book2018}. Note that these scaling features are well captured by the model of Grossmann and Lohse~\citep{Grossmann:JFM2000}; see also \citet{Pandey:PRE2016}.

In this paper, we perform DNS of 2D convection in a unit box for $Pr = 0.1$ and $Pr = 1$, for Rayleigh numbers between $10^6$ and $10^{12}$, not only for the purposes of exploring flow properties but also for highlighting the challenges of simulating 2D convection. We use the no-slip condition on all boundaries, with bottom and top walls isothermal and sidewalls adiabatic. In particular, we show that the velocity field evolves more slowly than the temperature, and call attention to large fluctuations that occur in what may be regarded as the nominal steady state of the velocity field: it is nominal in the sense that large-amplitude low-frequency oscillations persist around plausible averages. (To avoid excessive repetition, however, we often omit the qualifier ``nominal"). We relate these fluctuations to heat transport characteristics and provide estimates of suitably defined transient times. Heat transport also exhibits persistently wild fluctuations about its mean for strong thermal forcing. The main qualitative conclusion of this study is that such long transients, as well as wildly fluctuating heat transport, at least for aspect ratios of order unity, make the observation of the so-called ultimate state as elusive in 2D \citep{Doering:PRL2019} as in 3D \citep{Doering:PRL2020}.

\section{Numerical methodology} 
\label{sec:method}
We perform direct numerical simulations in a 2D fluid layer with horizontal dimension $L$, with an imposed temperature difference $\Delta T$ between the bottom and top plates separated by vertical dimension $H$. For this work, the aspect ratio $\Gamma = L/H = 1$. The following Oberbeck-Boussinesq equations dictating the flow are solved using {\sc Nek5000} solver \footnote{{\sc Nek5000} has been used extensively for the simulation of turbulent convection. For some details, see \cite{Scheel:NJP2013}}:
\begin{eqnarray}
\nabla \cdot {\bm u} & = & 0, \label{eq:m} \\ 
\frac{\partial {\bm u}}{\partial t} + {\bm u} \cdot \nabla {\bm u} & = & - \frac{ \nabla p}{\rho_0} + \alpha g (T-T_0) \hat{z} + \nu \nabla^2 {\bm u}, \label{eq:u} \\ 
\frac{\partial T}{\partial t} + {\bm u} \cdot \nabla T & = & \kappa \nabla^2 T. \label{eq:T}
\end{eqnarray}
Here, ${\bm u} = (u_x, u_z)$, $p$, and $T$ are the velocity, pressure, and temperature, respectively; $\rho_0$ is the reference density and $T_0$ is the reference temperature.
We non-dimensionalize equations~\eqref{eq:m}--\eqref{eq:T} using $H$, $\Delta T$, $u_f$, and $t_f$ as the scales for length, temperature, velocity, and time, respectively, where $u_f = \sqrt{\alpha g \Delta T H}$ is the free-fall velocity and $t_f = H/u_f$ is the free-fall time. The result contains the Prandtl number $Pr$ and the Rayleigh number $Ra = \alpha g \Delta T H^3/(\nu \kappa)$, where $\alpha$ is the isobaric thermal expansion coefficient and $g$ is the acceleration due to gravity.  
We explore two fluids with $Pr = 0.1$ and $Pr = 1$ for $Ra$ between $10^6$ and $10^{12}$. The square domain is decomposed into $N_e$ elements, and each element is further resolved using Lagrangian interpolation polynomials of order $N$ in both horizontal and vertical directions. Thus, the entire flow is resolved using $N_e N^2$ mesh cells. The no-slip condition for the velocity field is imposed on all boundaries. Isothermal and adiabatic conditions for the temperature field are imposed on the horizontal plates and sidewalls, respectively. 
The flow consists of a single large-scale structure covering the entire domain, except for $Pr = 1, Ra \leq 10^7$, where two vertically-stacked structures occur. The flow morphology is similar to that observed by \cite{Labarre:PRF2023} and \cite{Samuel:PRF2024}, so we do not show them here.

To resolve the thermal and viscous boundary layers, finer mesh is used near all boundaries. The spatial resolution in the flow is further ensured by computing the Kolmogorov length scale $\eta$, which is nominally the finest scale in the velocity field, and stipulating that the local vertical grid spacing $\Delta_z(z)$/$\eta(z) < 1.5$ for all simulations. The kinetic energy and the thermal dissipation rates are defined, respectively, as
\begin{eqnarray}
\varepsilon_u (\bm x) & = & \frac{\nu}{2} \sum_{l,m} \left( \frac{\partial u_l}{\partial x_m} + \frac{\partial u_m}{\partial x_l} \right)^2 \, , \\
\varepsilon_T (\bm x) & = & \kappa \left( \frac{\partial T}{\partial x_l} \right)^2 \, ,
\end{eqnarray}
and represent the rates of loss of kinetic energy and thermal energy per unit mass. Here, $l, m = (x,z)$, and the local Kolmogorov scale being given by $\eta = (\nu^3/\varepsilon_u)^{1/4}$. As the intermittent variation of $\varepsilon_u (\bm x)$ in the flow leads to variations in the Kolmogorov scale as well, we estimate an average
Kolmogorov scale in each horizontal plane using the horizontally- and temporally-averaged dissipation as
\begin{equation}
\eta(z) = \frac{\nu^{3/4}}{\langle \varepsilon_u \rangle_{x,t}^{1/4}(z)}.
\end{equation}
The finest length scale in the temperature field, the Batchelor scale $\eta/\sqrt{Pr}$, is of the same order as $\eta$, or coarser, in the present work. Thus, it is always adequately resolved. 


\section{Other associated definitions}
The Nusselt number in a horizontal plane is computed as~\citep{Chilla:EPJE2012}
\begin{equation}
Nu(z) = \frac{ \langle u_z T \rangle_{x,t} - \kappa \partial \langle T \rangle_{x,t}/\partial z} { \kappa \Delta T /H}, \label{eq:Nu_z}
\end{equation}
where $\langle u_z T \rangle_{x,t}$ is the convective component of the heat flux and $- \kappa \partial \langle T \rangle_{x,t}/\partial z$ is the diffusive component. As the vertical velocity vanishes at the horizontal plates, this relation yields the definition using the wall temperature gradient, as
\begin{equation}
Nu_W = - \frac{H}{\Delta T} \left. \frac{\partial \langle T \rangle_{x,t}} {\partial z} \right|_{z=0,H} \, . \label{eq:Nu_plate}
\end{equation}
Averaging equation~\eqref{eq:Nu_z} along the vertical direction yields the following relation for the global heat transport across the convective layer:
\begin{equation}
Nu = 1 + \frac{H}{\kappa \Delta T} \langle u_z T \rangle_{A,t} \, . \label{eq:Nu}
\end{equation}
Here $\langle \cdot \rangle_{A,t}$ stands for the average over the entire flow domain and simulation time covering the steady state.  

In another family of relations, the exact relations in RBC~\citep{Howard:ARFM1972, Shraiman:PRA1990} connect the Nusselt number $Nu$ with the globally-averaged kinetic energy dissipation rate $\varepsilon_u$ and the thermal dissipation rate $\varepsilon_T$, as
\begin{eqnarray}
\langle \varepsilon_u \rangle_{A,t} & = & \frac{\nu^3}{H^4} \frac{(Nu-1) Ra}{Pr^2} \label{eq:exact_u} \, , \\
\langle \varepsilon_T \rangle_{A,t} & = & \kappa \frac{(\Delta T)^2}{H^2} Nu. \label{eq:exact_T}
\end{eqnarray}
Thus, the Nusselt number can also be estimated from mean dissipation rates:
\begin{eqnarray}
Nu_{\varepsilon_u} & = & 1 + \frac{H^4}{\nu^3} \frac{Pr^2}{Ra} \langle \varepsilon_u \rangle_{A,t} \, , \label{eq:Nu_epsv} \\
Nu_{\varepsilon_T} & = & \frac{H^2}{(\Delta T)^2} \frac{\langle \varepsilon_T \rangle_{A,t}}{\kappa}. \label{eq:Nu_epst}
\end{eqnarray}
The agreement among simulated values from different definitions serves as a check on the accuracy and the adequacy of spatial and temporal resolutions~\citep{Stevens:JFM2010, Zhang:JFM2017, Pandey:PD2022}. In table~\ref{table:sim_detail}, we list the global heat flux estimated from all the methods and note that the agreement among various estimates of the Nusselt number is indeed very good. Also listed are other crucial parameters of the flow. We discuss below the approach to the final state of the Nusselt number using these definitions and relations, with comments on the scaling of global averages. In equations~\eqref{eq:Nu_z}--\eqref{eq:Nu} and \eqref{eq:Nu_epsv}--\eqref{eq:Nu_epst}, the Nusselt numbers are globally averaged quantities, though we do not show the averaging symbol explicitly, following standard usage of the past.

In the following, we study the evolutions of the instantaneous domain-averaged heat fluxes, which are defined as
\begin{eqnarray}
\overline{Nu} & = & 1 + \frac{H}{\kappa \Delta T} \langle u_z T \rangle_{A} \, , \label{eq:Nu_t} \\
\overline{Nu_{\varepsilon_u}} & = & 1 + \frac{H^4}{\nu^3} \frac{Pr^2}{Ra} \langle \varepsilon_u \rangle_{A} \, , \label{eq:Nu_epsv_t} \\
\overline{Nu_{\varepsilon_T}} & = & \frac{H^2}{(\Delta T)^2} \frac{\langle \varepsilon_T \rangle_{A}}{\kappa}. \label{eq:Nu_epst_t}
\end{eqnarray}

\begin{table}
\captionsetup{width=1\textwidth}
  \begin{center}
\def~{\hphantom{0}}
  \begin{tabular}{lccccccccc}
$Pr$ & $Ra$ & $N_e N^2$ & $Nu$ & $Nu_{\varepsilon_u}$ & $Nu_{\varepsilon_T}$ &  $Nu_W$ & $Re$ & $t_{sim} \, (t_f)$ 
\vspace{2mm} 
\\
0.1 &         $10^{6}$ &   $690^2$ 	 & 	  5.91 $\pm$    0.1 &   5.91 $\pm$    0.4 &   5.91 $\pm$    0.4 &   5.91 $\pm$    0.5 &   1916 $\pm$     96 &    605 \\ 
0.1 & $2\times 10^{6}$ &   $300^2$ 	 & 	  7.05 $\pm$    0.1 &   7.05 $\pm$    0.1 &   7.05 $\pm$    0.1 &   7.05 $\pm$    0.9 &   2747 $\pm$     18 &   1159 \\ 
0.1 & $3\times 10^{6}$ &   $300^2$ 	 & 	  7.88 $\pm$    0.1 &   7.88 $\pm$    0.1 &   7.88 $\pm$    0.1 &   7.88 $\pm$    1.2 &   3396 $\pm$     27 &   1184 \\ 
0.1 & $6\times 10^{6}$ &   $300^2$ 	 & 	 10.04 $\pm$    0.7 &   9.97 $\pm$    0.6 &   9.97 $\pm$    1.2 &   9.98 $\pm$    1.6 &   4979 $\pm$    428 &   2684 \\ 
0.1 &         $10^{7}$ &   $690^2$ 	 & 	 11.77 $\pm$    1.1 &  11.78 $\pm$    0.9 &  11.75 $\pm$    1.3 &  11.73 $\pm$    1.7 &   6521 $\pm$    514 &   2152 \\ 
0.1 & $2\times 10^{7}$ &   $500^2$ 	 & 	 15.35 $\pm$    2.1 &  15.32 $\pm$    1.1 &  15.29 $\pm$    1.5 &  15.30 $\pm$    1.9 &   9512 $\pm$    968 &   3007 \\ 
0.1 & $3\times 10^{7}$ &   $690^2$ 	 & 	 17.50 $\pm$    8.6 &  17.52 $\pm$    3.2 &  17.52 $\pm$    2.5 &  17.48 $\pm$    3.0 &  12065 $\pm$   2981 &   1415 \\ 
0.1 &         $10^{8}$ &  $1150^2$ 	 & 	 23.01 $\pm$     12 &  22.97 $\pm$    4.3 &  23.05 $\pm$    3.0 &  23.06 $\pm$    3.7 &  22316 $\pm$   5024 &   1081 \\ 
0.1 & $3\times 10^{8}$ &  $1150^2$ 	 & 	 31.06 $\pm$     16 &  30.82 $\pm$    7.2 &  30.75 $\pm$    3.7 &  30.76 $\pm$    4.7 &  43375 $\pm$   9299 &   1116 \\ 
0.1 &         $10^{9}$ &  $2070^2$ 	 & 	 42.72 $\pm$     24 &  42.28 $\pm$     12 &  42.70 $\pm$    4.9 &  42.72 $\pm$    6.5 &  95499 $\pm$  15598 &    304 \\ 
0.1 & $3\times 10^{9}$ &  $2230^2$ 	 & 	 59.15 $\pm$     41 &  58.71 $\pm$     14 &  59.04 $\pm$    6.2 &  58.99 $\pm$    7.8 & 200001 $\pm$  22805 &    222 \\ 
0.1 &        $10^{10}$ &  $3122^2$ 	 & 	 84.74 $\pm$     58 &  79.24 $\pm$     26 &  84.94 $\pm$    8.6 &  84.91 $\pm$     12 & 446638 $\pm$  56213 &    178 \\ 
  1 & $3\times 10^{6}$ &   $300^2$ 	 & 	  6.01 $\pm$    0.3 &   6.01 $\pm$    0.1 &   6.01 $\pm$    0.1 &   6.01 $\pm$    0.6 &    289 $\pm$     33 &   6375 \\ 
  1 & $6\times 10^{6}$ &   $300^2$ 	 & 	  7.11 $\pm$    0.1 &   7.11 $\pm$    0.1 &   7.11 $\pm$    0.1 &   7.11 $\pm$    0.3 &    423 $\pm$      4 &   6623 \\ 
  1 & $8\times 10^{6}$ &   $300^2$ 	 & 	  8.18 $\pm$    0.1 &   8.18 $\pm$    0.1 &   8.18 $\pm$    0.2 &   8.18 $\pm$    0.7 &    530 $\pm$     21 &   3630 \\ 
  1 & $9\times 10^{6}$ &   $300^2$ 	 & 	  8.64 $\pm$    0.2 &   8.64 $\pm$    0.1 &   8.65 $\pm$    0.2 &   8.65 $\pm$    0.8 &    580 $\pm$     21 &   5512 \\ 
  1 &         $10^{7}$ &   $300^2$ 	 & 	  9.03 $\pm$    0.2 &   9.03 $\pm$    0.1 &   9.03 $\pm$    0.2 &   9.03 $\pm$    1.0 &    626 $\pm$     25 &   5188 \\ 
  1 & $1.5\times 10^{7}$ &   $300^2$ 	 & 	 13.33 $\pm$    1.2 &  13.35 $\pm$    0.7 &  13.34 $\pm$    1.2 &  13.33 $\pm$    1.7 &    918 $\pm$    129 &   7056 \\ 
  1 & $2\times 10^{7}$ &   $690^2$ 	 & 	 14.75 $\pm$    4.5 &  14.74 $\pm$    1.2 &  14.83 $\pm$    1.7 &  14.79 $\pm$    1.8 &   1108 $\pm$    366 &   1879 \\ 
  1 & $3\times 10^{7}$ &   $690^2$ 	 & 	 17.27 $\pm$    3.4 &  17.00 $\pm$    1.4 &  16.98 $\pm$    1.9 &  17.10 $\pm$    2.1 &   1376 $\pm$    337 &   2782 \\ 
  1 & $6\times 10^{7}$ &   $690^2$ 	 & 	 21.48 $\pm$    4.8 &  21.57 $\pm$    1.8 &  21.57 $\pm$    2.3 &  21.56 $\pm$    2.4 &   2010 $\pm$    442 &   2584 \\ 
  1 &         $10^{8}$ &   $690^2$ 	 & 	 25.01 $\pm$    5.6 &  25.20 $\pm$    2.0 &  25.22 $\pm$    2.5 &  25.21 $\pm$    2.3 &   2665 $\pm$    432 &   1717 \\ 
  1 & $3\times 10^{8}$ &   $690^2$ 	 & 	 35.61 $\pm$    7.5 &  35.85 $\pm$    1.8 &  35.84 $\pm$    2.2 &  35.88 $\pm$    2.0 &   4936 $\pm$    627 &   2379 \\ 
  1 & $6\times 10^{8}$ &   $690^2$ 	 & 	 44.15 $\pm$     12 &  44.01 $\pm$    2.3 &  44.04 $\pm$    2.2 &  44.02 $\pm$    2.2 &   7385 $\pm$    954 &   1178 \\ 
  1 &         $10^{9}$ &   $700^2$ 	 & 	 50.87 $\pm$     35 &  51.14 $\pm$     10 &  51.29 $\pm$    5.0 &  51.20 $\pm$    6.3 &  12012 $\pm$   3324 &   2112 \\ 
  1 & $3\times 10^{9}$ &  $1150^2$ 	 & 	 67.15 $\pm$     43 &  68.32 $\pm$     12 &  68.17 $\pm$    6.3 &  68.26 $\pm$    8.4 &  20958 $\pm$   5152 &   1479 \\ 
  1 &        $10^{10}$ &  $1150^2$ 	 & 	 94.52 $\pm$     57 &  94.93 $\pm$     21 &  95.26 $\pm$    8.5 &  94.97 $\pm$     12 &  43533 $\pm$   8819 &    989 \\ 
  1 & $3\times 10^{10}$ &  $1610^2$ 	 & 	 130.2 $\pm$     84 &  129.7 $\pm$     34 &  130.3 $\pm$     12 &  130.3 $\pm$     18 &  89031 $\pm$  18989 &   1105 \\ 
  1 &        $10^{11}$ &  $2070^2$ 	 & 	 183.3 $\pm$    128 &  184.1 $\pm$     56 &  185.2 $\pm$     17 &  184.7 $\pm$     23 & 199858 $\pm$  41998 &   1268 \\ 
  1 & $3\times 10^{11}$ &  $3122^2$ 	 & 	 262.3 $\pm$    185 &  260.7 $\pm$    109 &  262.2 $\pm$     24 &  262.1 $\pm$     36 & 395763 $\pm$  69299 &    336 \\ 
  1 &        $10^{12}$ &  $5210^2$ 	 & 	 394.2 $\pm$    286 &  376.1 $\pm$     88 &  395.8 $\pm$     36 &  395.3 $\pm$     54 & 1085094 $\pm$  54592 &    253 \\ 
  \end{tabular}
  \caption{Important parameters of direct numerical simulations in a 2D box of $\Gamma = 1$. We list the Prandtl number, the Rayleigh number, the total number of mesh cells in the entire flow domain $N_e N^2$, the Nusselt number using equations \eqref{eq:Nu}, \eqref{eq:Nu_epsv}, \eqref{eq:Nu_epst}, and \eqref{eq:Nu_plate}, respectively, the Reynolds number, and the simulation time after the flow attains a steady state, $t_{sim}$. Error bars in Nusselt and Reynolds numbers are the corresponding standard deviations.}  
  \label{table:sim_detail}
  \end{center}
\end{table}

\section{Transient characteristics}
\label{sec:transient}

\begin{figure}
\captionsetup{width=1\textwidth}
\centerline{
\includegraphics[width=1\textwidth]{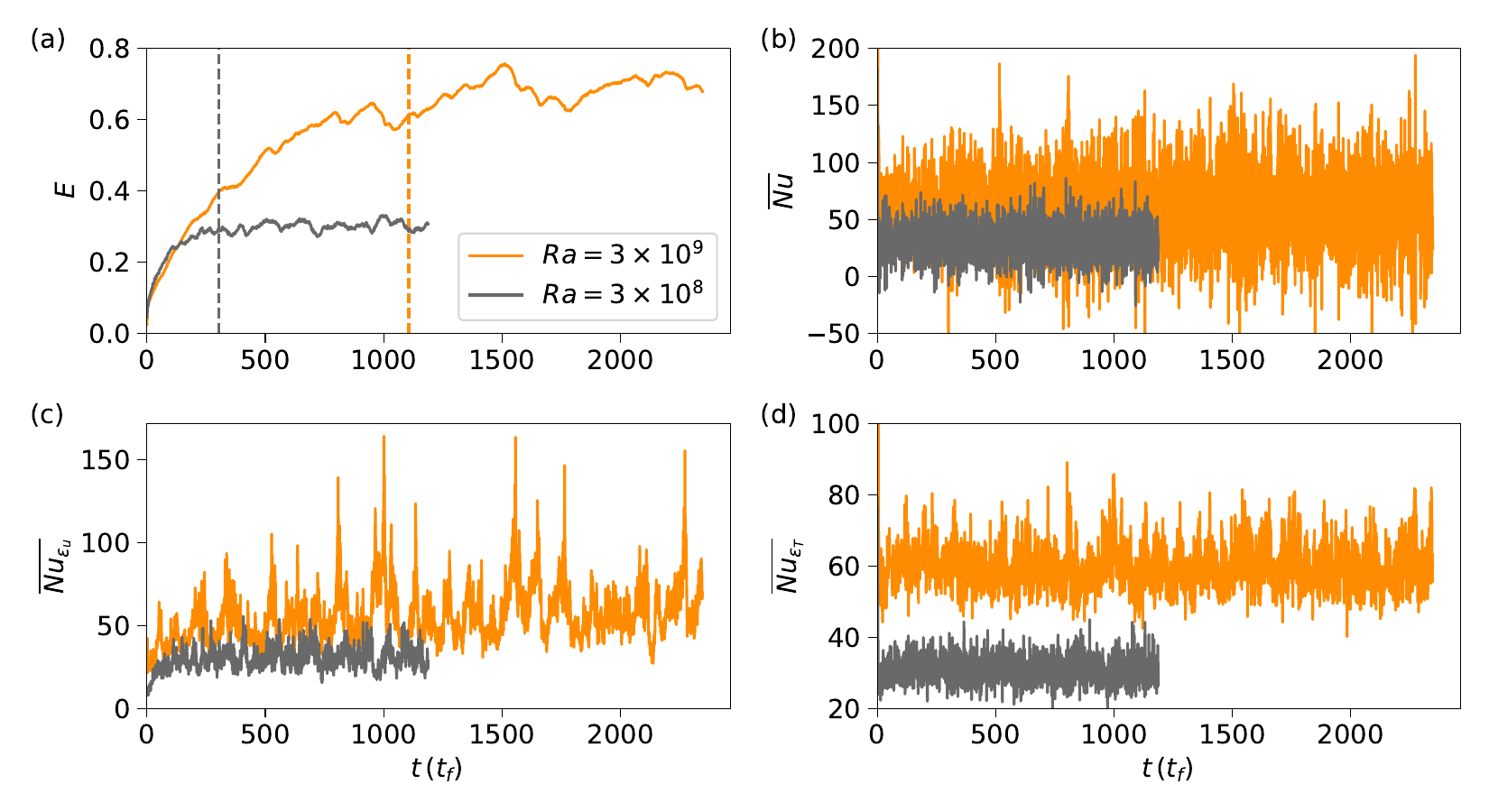}}
\caption{Evolution of the integral quantities in the transient state for $Pr = 0.1$, and $Ra = 3 \times 10^8$ (black curves) and $Ra = 3 \times 10^9$ (orange curves). (a) The domain-averaged kinetic energy $E$ increases slowly and takes a few thousand free-fall times to reach the steady state. The transient time is longer for higher $Ra$. Dashed vertical lines show a quantitative measure of the transient time $t_{trns}$, obtained from equation~\eqref{eq:t_trans}, to be discussed later. Panels (b), (c) and (d) show that the Nusselt number fluctuates rapidly about its mean nearly from the start, but fluctuations have different characters depending on the definition of the Nusselt number. In (b), the fluctuations in $\overline{Nu}$ are strong and of high frequency with no well-defined transient state, and there is an overlap for the two $Ra$. (c) shows that one can roughly identify a transient state in $\overline{Nu_{\varepsilon_u}}$, which exhibits very strong fluctuations containing both high and low frequencies. (d) shows that $\overline{Nu_{\varepsilon_T}}$ fluctuates similar to $\overline{Nu}$ in (b), but there is no overlap for the two Rayleigh numbers. The occasional appearance of negative values of $\overline {Nu}$ suggests the likelihood that a small parcel of the coldest fluid from the top wall registers directly at the bottom wall, and vice versa.}
\label{fig:integ_t_trans}
\end{figure}

A common method for initiating high-$Ra$ simulations is to start them from the flow at a lower $Ra$. Simulations can also be performed {\it ab initio} from the conduction state with random perturbations. In both cases, global heat flux and kinetic energy evolve with time, and one needs to wait some time before a statistically steady state is attained. Figure~\ref{fig:integ_t_trans} shows the temporal evolution of integral quantities for the two $Ra$ indicated and $Pr = 0.1$, with simulations initiated from the conduction state. On the one hand, we observe that $\overline{Nu}$ and $\overline{Nu_{\varepsilon_T}}$ in figure~\ref{fig:integ_t_trans}(b,d) start oscillating about some mean value shortly after the simulation begins. On the other hand, $\overline{Nu_{\varepsilon_u}}$ in figure~\ref{fig:integ_t_trans}(c) initially increases and starts to fluctuate about a mean value only after some time has elapsed. The instantaneous domain-averaged non-dimensional kinetic energy $E = \langle (u_x^2 + u_z^2)/2 \rangle_A/u_f^2$ in figure~\ref{fig:integ_t_trans}(a) offers the best means to determine the time to the steady state. There is no ambiguity about this approach for the lower $Ra$; however, aside from the fact that it takes longer at the higher $Ra$ for the energy to achieve its steady state, the latter is somewhat nominal because fluctuations about the average are significant. The situation becomes more so at even higher Rayleigh numbers. Fluctuations in the domain-averaged energy $E$ follow from dynamical considerations; we shall show this in \S~\ref{sec:balance}.

\begin{figure}
\captionsetup{width=1\textwidth}
\centerline{
\includegraphics[width=0.6\textwidth]{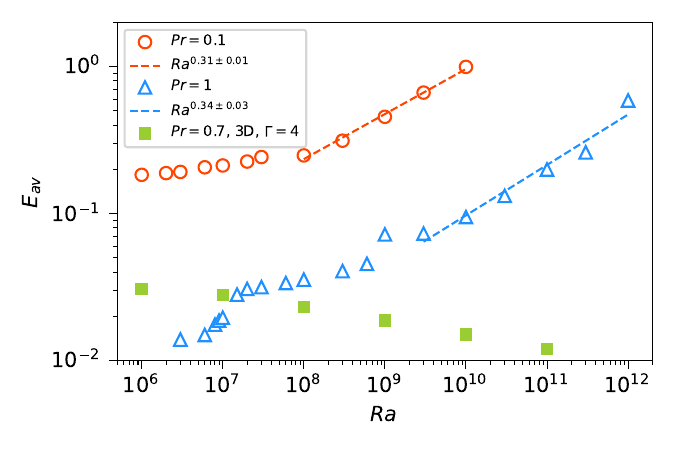}}
\caption{Global (area- as well as time-averaged) kinetic energy in the steady state $E_{av}$ as a function of $Ra$. An increasing trend with $Ra$ is observed in 2D RBC. In the turbulent regimes ($Ra \geq 10^8$ for $Pr = 0.1$ and $Ra > 10^{9}$ for $Pr = 1$), the data roughly follow $E_{av} \sim Ra^{1/3}$, shown as dashed lines. In contrast, $E_{av}$ in 3D RBC for $Pr = 0.7$ (taken from \citet{Samuel:JFM2024}) shows a weakly decreasing trend. The transitions observed in this figure are related to transitions in the flow structure; see \S~\ref{sec:Trms}. The statistical error bars in almost all the figures here are comparable to the thickness of the symbols. The exception is figure 6 for which the errors bars are shown explicitly.}
\label{fig:E}
\end{figure}

We define the average kinetic energy in the steady state, $E_{av}$, as
\begin{equation}
E_{av} = \frac{1}{2}u_{RMS}^2/u_f^2 \, ,
\end{equation}
where $u_{RMS} = \sqrt{ \langle u_x^2 + u_z^2 \rangle_{A,t}}$ is the root mean square (RMS) velocity of the flow in the steady state, and plot it in figure~\ref{fig:E} as a function of $Ra$. Before discussing differences between the two Prandtl numbers, we note the major difference between 2D and 3D fluctuations. In the 3D case, for low and moderate Prandtl numbers, $E_{av}$ is a slowly decreasing function of $Ra$; for example, $E_{av}$ from a horizontally periodic cuboid of $\Gamma = 4$ for $Pr = 0.7$ (from \citet{Samuel:JFM2024}), shown as green squares in figure~\ref{fig:E}, follows a $Ra^{-0.07}$ scaling. The corresponding behavior for 2D cases is that the fluctuations increase with $Ra$. The behavior is different for different $Pr$ when the Rayleigh numbers are low, but follows a roughly $Ra^{1/3}$ scaling at high Rayleigh numbers, as indicated by the blue and red dashed lines in figure~\ref{fig:E}.\footnote{Our objective is not a detailed study of differences between 2D and 3D convection. Some such differences have been pointed out, e.g., by \citet{Poel:JFM2013} and \citet{Pandey:Pramana2016}.}  The challenges posed by this increasing trend will be discussed below.

\begin{figure}
\captionsetup{width=1\textwidth}
\centerline{
\includegraphics[width=0.6\textwidth]{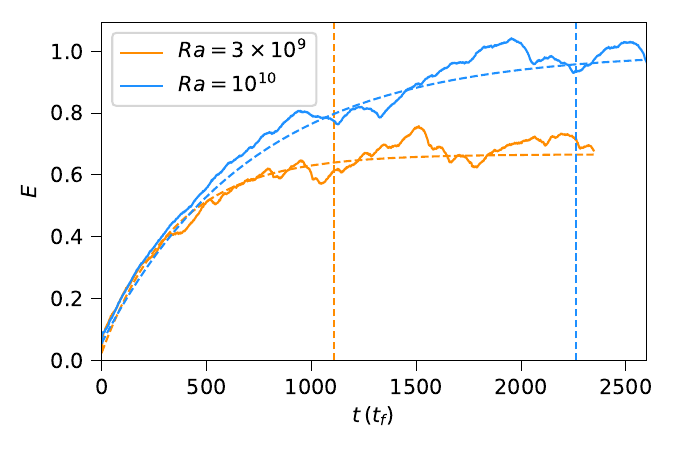}}
\caption{Evolution of the domain-averaged kinetic energy $E(t)$ in the transient state for $Pr = 0.1$ and $Ra = 3 \times 10^9$ and $Ra = 10^{10}$ can be described well by equation~\eqref{eq:E_growth} and the dashed curves show $E_\mathrm{fit}(t)$ for the growth rate. Vertical lines show that the transient time $t_{trns}$ is much longer for $Ra = 10^{10}$ than for $Ra = 3 \times 10^9$.}
\label{fig:deter_trans}
\end{figure}

It is useful to study the variation of $t_{trns}$ (which is the time required for $E$ to attain some stipulated fraction of $E_{av}$) with respect to $Ra$ and $Pr$ by some simple scheme. To this end, we plot the evolution of $E$ for $Ra = 3 \times 10^9$ and $Ra = 10^{10}$ for $Pr = 0.1$ in figure~\ref{fig:deter_trans}. The growth of $E$ for both $Ra$ is comparable for initial times but the curve for higher $Ra$ continues to grow for a longer period before a nominally steady state is reached, but with conspicuous fluctuations. 

Figure~\ref{fig:deter_trans} suggests that $E(t)$ can be fitted with an equation of the form
\begin{equation}
[E_{av} - E(t)]/E_{av} = c \exp(-kt), \label{eq:E_growth}
\end{equation}
where $E_{av}$ is the ``steady-state" or ``asymptotic" mean energy of the flow; the growth rate $k$ depends on Rayleigh and Prandtl numbers. The factor $c$ captures the finite energy at the initial instant in figure~\ref{fig:deter_trans}: though $E(t)$ grows rapidly only when the convective motion is established, there is a finite $E(t)$ from which it starts.  

\begin{figure}
\captionsetup{width=1\textwidth}
\centerline{
\includegraphics[width=1\textwidth]{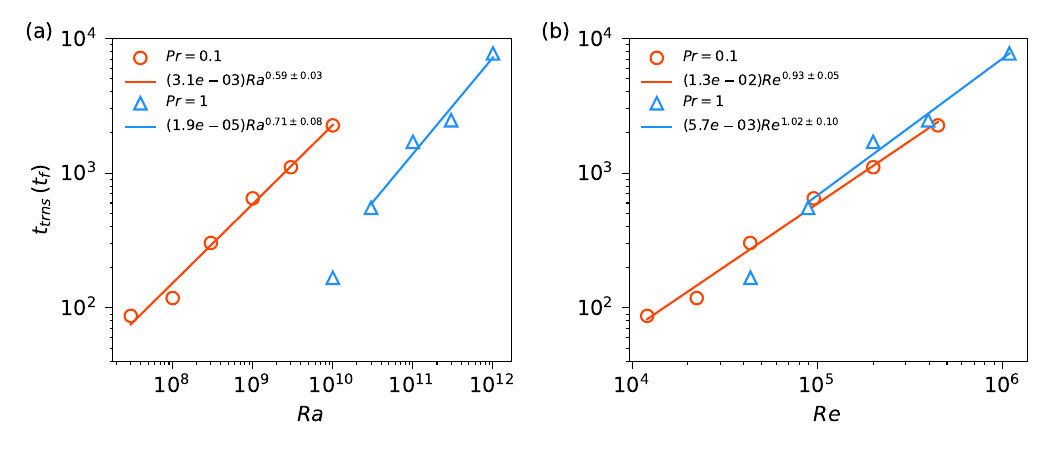}}
\caption{(a) The transient time $t_{trns}$ as a function of $Ra$ shows a power law. For a given $Ra$, $t_{trns}$ is longer for lower $Pr$. (b) The $t_{trns}$ as a function of $Re$ is essentially the same for both values of $Pr$ and exhibits a nearly linear trend. Moreover, $t_{trns}$ is nearly the same for both $Pr$ when the Reynolds numbers are the same. Data correspond only to the turbulent regimes. Sparser data sets for $Pr = 0.021$ are consistent with the Prandtl numbers of these figures.}
\label{fig:t_trns}
\end{figure}

We fit the suitable segment of the growth curve $E(t)$ with equation~\eqref{eq:E_growth}. The segment at late times is not suitable for fitting the formula because the kinetic energy does not attain a constant value but fluctuates strongly for long periods of time. The scaling factor $c$ is calculated from the data as $c = (E_{av} - E(0))/E_{av}$, where $E(0)$ is the energy at $t = 0$. The dashed curves in figure~\ref{fig:deter_trans} are fits to the data. Having determined the growth rate, we define the transient time $t_{trns}$ as the time when the fitted curve $E_{fit}(t)$ in figure~\ref{fig:deter_trans} reaches 96\% of its `constant' value, $E_{av}$. Thus, the transient time is estimated as
\begin{equation}
t_{trns} = \frac{1}{k} \log \frac{c}{0.04} \, \label{eq:t_trans}.
\end{equation}
The transient time thus estimated is plotted as a function of $Ra$ in figure~\ref{fig:t_trns}(a). Note that $t_{trns}$ increases as a power-law; for $Pr = 0.1$, the best fit yields $t_{trns} \sim Ra^{0.59 \pm 0.03}$, while we find $t_{trns} \sim Ra^{0.71 \pm 0.08}$ for $Pr = 1$.

We have experimented with different definitions of $t_{trns}$ (e.g., by requiring it to reach 90\% of $E_{av}$) and find the same trends, though the numbers are different. The fact that the {$Ra$-exponent} is non-trivial suggests an important role for the boundary layers and their relation to the large structure; those details (as well as the effect of the aspect ratio) are in need for further exploration.

\begin{figure}
\captionsetup{width=1\textwidth}
\centerline{
\includegraphics[width=1\textwidth]{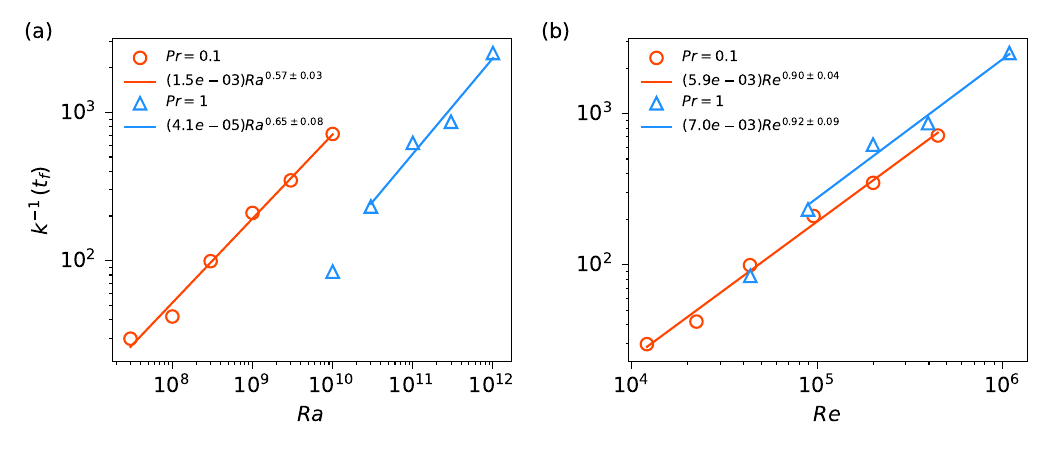}}
\caption{Inverse growth rate as a function of (a) $Ra$ and (b) $Re$. The trends with $Re$ are approximately the same for both Prandtl numbers, and are slightly below linear.}
\label{fig:growth}
\end{figure}

Figure~\ref{fig:t_trns}(a) further shows that for a given $Ra$, $t_{trns}$ is longer for $Pr = 0.1$ than for $Pr = 1$. This is expected as $u_{RMS}$---consequently the Reynolds number---is higher for smaller $Pr$ \citep{Schumacher:PNAS2015, Pandey:PoF2016, Pandey:PD2022}. In figure~\ref{fig:t_trns}(b), we plot $t_{trns}$ as a function of the Reynolds number $Re$, which is defined as
\begin{equation}
Re = u_{RMS}H/\nu \, . \label{eq:Re}
\end{equation}
We find that the two $t_{trns}$ fall approximately on the same line for both Prandtl numbers, the best fit for which is nearly linear. Thus, a 2D RBC at high-$Ra$ has to be simulated for very long times to achieve a statistically steady state. If a steady state is not achieved and $E$ is still growing, one will find $Nu_{\varepsilon_u} < Nu$ (see \S~\ref{sec:balance}) even when spatial and temporal resolutions are adequate.

Shown in figure~\ref{fig:growth} is the variation of the inverse growth rate $k^{-1}$ with respect to $Ra$ (figure~\ref{fig:growth}(a)) and $Re$ (figure~\ref{fig:growth}(b)). It appears that $k^{-1}$ is approximately a linear function of $Re$ and depends only weakly on the Prandtl number.

A longer transient at higher $Ra$ compels researchers to use some workaround to achieve the statistically steady state in a shorter time. For example, one might start with a much coarser resolution and ramp it up to the required level only after the kinetic energy has reached an approximate steady state. It is practically impossible with available computing power to conduct very high-$Ra$ simulation with fully-resolved fields in the entire transient state. One can use a coarser mesh during the transient when all degrees of freedom have not yet been excited. However, it is important to ensure that simulations on the coarser mesh lead to the same steady state as that attained in a well-resolved simulation. A concern otherwise would be that the coarse-grid simulation results in different kinetic energy from the properly resolved case, ultimately affecting the scaling of $Re$, particularly because of the presence of the low frequency modes (see \S~\ref{sec:balance}). Though we demonstrate in Appendix~\ref{sec:apndx} that the mean kinetic energy in the steady state and the transient time are essentially the same in coarser simulations, it is not necessarily likely to be a general result. The data in figures~\ref{fig:integ_t_trans} and \ref{fig:deter_trans} correspond to simulations at the coarser resolution. 

\section{Scaling exponents for integral transport in the steady state}
\label{sec:scaling}

\subsection{Heat transport}
\label{sec:Nu}

\begin{figure}
\captionsetup{width=1\textwidth}
\centerline{
\includegraphics[width=1\textwidth]{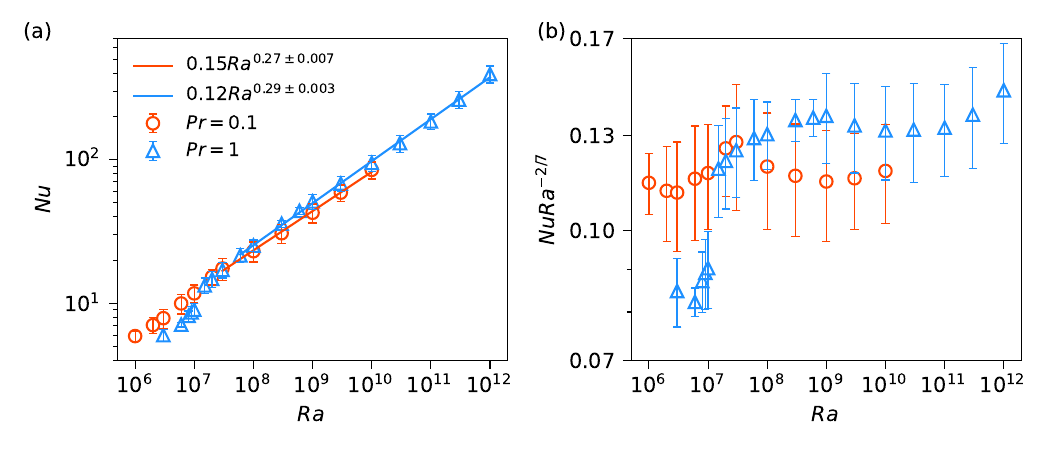}}
\caption{(a) Nusselt number as a function of $Ra$ for $Pr = 0.1$ (red circles) and $Pr = 1$ (blue triangles). The scaling for high Rayleigh numbers differs only slightly from the low-Ra behavior. (b) Normalized Nusselt number $Nu Ra^{-2/7}$ shows that the $2/7$-ths scaling is only approximate for moderate Rayleigh numbers and moderate aspect ratios. {Wall heat flux from equation~\eqref{eq:Nu_plate} is shown here with the error bars representing the standard deviation.}}
\label{fig:nu}
\end{figure}

We plot $Nu$, computed from equation~\eqref{eq:Nu_plate}, as a function of $Ra$ in figure~\ref{fig:nu}(a). The data for high Rayleigh numbers seem to closely follow similar power laws for both $Pr$. For $Pr = 1$, the best fit for $Ra \geq 6 \times 10^7$ yields the scaling $Nu = 0.12 Ra^{0.29 \pm 0.003}$. The exponent is close to $2/7$ proposed for the so-called `hard turbulence' in confined 3D RBC~\citep{Castaing:JFM1989}, also explored in various later studies~\citep{Siggia:ARFM1994, Chilla:EPJE2012, Johnston:PRL2009}. We plot the normalized Nusselt number $Nu Ra^{-2/7}$ versus $Ra$ in figure~\ref{fig:nu}(b), which reveals that the local exponent varies considerably and is only approximately $2/7$. Furthermore, we note that the Nusselt numbers fluctuate significantly, being significantly larger when $Nu$ is computed using equation~\eqref{eq:Nu}. To some extent, this result indicates that the local scaling exponents obtained by fitting heat transport data over short ranges of Rayleigh numbers could lead to misleading conclusions. Figure~\ref{fig:nu}(a) also shows that the Nusselt numbers for $Ra \leq 10^7$ do not follow the higher-$Ra$ trend. This is a consequence of differing flow structures observed: the flow for $Ra \leq 10^7$ consists of two vertically-stacked rolls, whereas, for $Ra > 10^7$, a single-roll structure is observed. These findings suggest that the flow is less efficient in transporting heat when the double-roll state occurs, which is in line with previous observations in both 2D and 3D~\citep{Xi:POF2008, Weiss:JFM2011, Poel:PRE2011}. 

For $Ra \leq 3 \times 10^7$, the Nusselt numbers for $Pr = 0.1$ are higher than those for $Pr = 1$. Unlike for $Pr = 1$, the flow for $Pr = 0.1$ shows no double-roll state at lower Rayleigh numbers, which is why a larger difference appears in $Nu$ for the two Prandtl numbers. Figure~\ref{fig:nu}(b) clearly shows that heat transport is smaller for $Pr = 0.1$ than for $Pr = 1$ in the turbulent regime, i.e., for $Ra > 3 \times 10^7$. This feature is similar to that observed in 3D RBC, where lower-$Pr$ fluids are less efficient at transporting heat when $Pr < 1$~\citep{Verzicco:JFM1999, Poel:JFM2013, Pandey:EPL2021, Pandey:JFM2022}. The Grossmann-Lohse model~\citep{Grossmann:JFM2000, Grossmann:PRL2001} also suggests a similar trend. 

\subsection{Momentum transport}
\label{sec:Re}

\begin{figure}
\captionsetup{width=1\textwidth}
\centerline{
\includegraphics[width=0.6\textwidth]{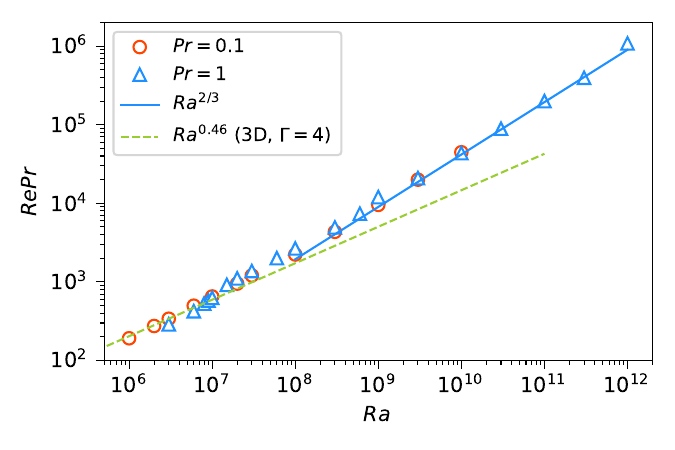}}
\caption{The Reynolds number based on $u_{RMS}$ scales nearly as $Ra^{2/3}$ in the turbulent regime (indicated by the blue solid line), which is distinctively different from scaling $Ra^{1/2}$ reported in 3D RBC. Green dashed line indicates the $Re \sim Ra^{0.46}$ scaling observed for 3D RBC in a $\Gamma = 4$ box by \citet{Samuel:JFM2024}.}
\label{fig:Re}
\end{figure}

A variety of velocity scales can be defined in turbulent RBC, and can be used to define the Reynolds number $Re$. In cylindrical or cubic domains with $\Gamma \approx 1$, the most dominant eddy in the flow is in the form of a large-scale circulation, and its velocity is observed to scale with the free-fall velocity~\citep{Lam:PRE2002, Xia:PRE2003}. In DNS, the Reynolds number is often obtained using the RMS velocity and the depth of the convective layer $H$ (see equation~\eqref{eq:Re})~\citep{Scheel:JFM2016, Pandey:JFM2022}. We plot $RePr$ against $Ra$ in figure~\ref{fig:Re}. 
Also indicated by the dashed line is the powerlaw $RePr = 0.33 Ra^{0.46}$, as found by \citet{Samuel:JFM2024} for $Pr = 0.7$. There is some overlap in the magnitude of the Reynolds numbers between 2D and 3D RBC, but the scaling exponents differ, especially for high Rayleigh numbers. In the high-$Ra$ regime, we find that the Reynolds number scales as $Re = 0.13 Ra^{0.65 \pm 0.01}$ for $Pr = 0.1$ and $Re = 0.02 Ra^{0.65 \pm 0.01}$ for $Pr = 1$. The data for lower Rayleigh numbers exhibit approximately the $Re \sim Ra^{0.55}$ scaling for both $Pr$. 

As is well known, the Reynolds number decreases with increasing $Pr$ in 3D RBC~\citep{Poel:JFM2013, Pandey:EPL2021}; similarly, in 2D, the Reynolds number scales approximately as $Ra^{2/3}Pr^{-1}$, consistent with the result found semi-analytically by \citet{Chini:PoF2009, Wen:JFM2020}. This $Re$ scaling further suggests that 
\begin{equation}
\frac{u_{RMS}}{u_f} = Re \sqrt{\frac{Pr}{Ra}} \sim Ra^{1/6} Pr^{-1/2} \, , \label{eq:urms}
\end{equation}
and consequently 
\begin{equation}
E_{av} = \frac{1}{2} \frac{u^2_{RMS}}{u^2_f} \sim Ra^{1/3} Pr^{-1}. \label{eq:E}
\end{equation}
The high-$Ra$ data in figure~\ref{fig:E} are indeed consistent with this scaling.

\subsection{RMS temperature fluctuation}
\label{sec:Trms}

\begin{figure}
\captionsetup{width=1\textwidth}
\centerline{
\includegraphics[width=1\textwidth]{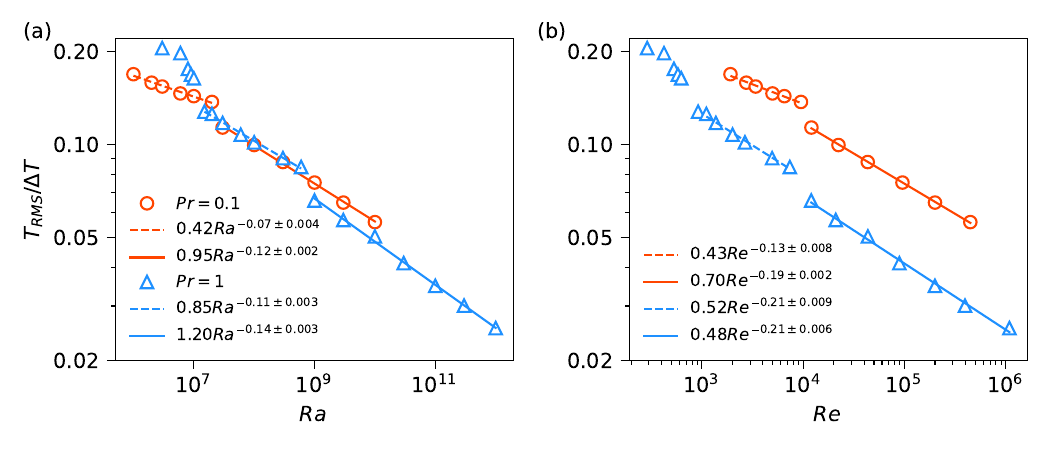}}
\caption{(a) Root mean square temperature fluctuation $T_{RMS}$ decreases with $Ra$ for both $Pr$ and have nearly the same magnitude for moderate Rayleigh numbers. Different exponents as well as prefactors are found for moderate and high Rayleigh numbers. (b) $T_{RMS}$ as a function of $Re$ shows that it scales approximately as $Re^{-0.2}$ in the turbulent regime (for $Re > 10^4$) for both Prandtl numbers. }
\label{fig:Trms}
\end{figure}

We compute the RMS temperature fluctuation as
\begin{equation}
T_{RMS} = \sqrt{ \langle T^2 \rangle_{A,t} - \langle T \rangle_{A,t}^2}
\end{equation}
and plot $T_{RMS}/\Delta T$ as a function of $Ra$ in figure~\ref{fig:Trms}(a). It is clear that $T_{RMS}/\Delta T$ decreases with increasing $Ra$ but various regions can be identified in figure~\ref{fig:Trms}(a). For $Pr = 0.1$, $T_{RMS}$ for $Ra \leq 2 \times 10^7$ exhibits $Ra^{-0.07}$ scaling but it starts, somewhat abruptly, to follow a steeper $Ra^{-0.12}$ scaling for higher $Ra$. For $Pr = 1$, too, we find that $T_{RMS}$ for $Ra < 10^9$ shows a scaling of $Ra^{-0.11}$, while a scaling of $Ra^{-0.14}$ ensues for $Ra \geq 10^9$. Again, the transition between these two regimes is nearly abrupt. The RMS fluctuations for $Ra \leq 10^7$ are larger and clearly depart from the trend for higher $Ra$; this occurs because of the double-roll state observed for weak thermal forcing at $Pr = 1$. It is interesting that the scaling $Ra^{-0.14}$ for high $Ra$ agrees well with those of fluctuations at the center of cylindrical RBC cells for $Pr \approx 0.7$~\citep{Castaing:JFM1989, Niemela:Nature2000}, as well as with that in the bulk region of a horizontally-periodic box for $Pr = 0.7$, $\Gamma = 4$~\citep{Samuel:JFM2024}; see also \citet{Pandey:PoF2016}. We also note that the magnitude of $T_{RMS}$ for moderate Rayleigh numbers ($10^7 < Ra < 10^9$) is quite similar for the two Prandtl numbers. 

The nearly abrupt transitions in $T_{RMS}$ are related to changes in flow morphology and statistics, and in the corresponding heat transport scaling. Data for $Pr = 0.1$ show that the scaling exponent $\gamma$ in the $Nu$-$Ra$ relation changes from $\approx 0.35$ to $\approx 0.25$ for $Ra \geq 3 \times 10^7$, with the latter value persisting for nearly two decades of $Ra$ (i.e., up to $Ra \approx 10^9$), beyond which it again increases to 0.30.  Similarly, for $Pr = 1$, $\gamma$ changes from $\approx 0.31$ to $\approx 2/7$ after the transition at $Ra \approx 10^9$. \citet{Labarre:PRF2023} recently noted that the ratio between the RMS fluctuations and the mean heat flux increases abruptly in 2D once $Ra/Pr$ increases beyond $\approx 10^9$. A somewhat similar transition was reported also by \citet{Gao:JFM2024}, who found the transition $Ra$ to scale with the Prandtl number as $Pr^{1.41}$. Our own data show that the relative fluctuation in the heat flux near the wall and in the bulk are enhanced significantly for $Ra \geq 3 \times 10^7$ at $Pr = 0.1$ and for $Ra \geq 10^9$ at $Pr = 1$ (see table~\ref{table:sim_detail}).

The decrease in $T_{RMS}/\Delta T$ with $Ra$ is related to the thermal boundary layer, which becomes thinner with $Ra$~\citep{Pandey:JFM2021, Scheel:JFM2012}. As $T_{RMS}$ represents an average measure of the temperature anomaly in the flow, the dominant contribution to $T_{RMS}$ arises from regions occupied by thermal plumes. This is because the temperature within the plumes varies slowly and differs strongly from the ambient temperature, which is approximately the mean temperature $\Delta T/2$ in the flow. As the fraction of the volume occupied by the plumes decreases with $Ra$, so does their contribution to $T_{RMS}/\Delta T$. A similar magnitude of RMS fluctuations for the two Prandtl numbers (and moderate Rayleigh numbers) is due to the similar nature in the two cases of heat transport (see figure~\ref{fig:nu}), which determines the thickness of the thermal boundary layer. 

In figure~\ref{fig:Trms}(b), we show $T_{RMS}/\Delta T$ as a function of $Re$. Although the data for the two Prandtl numbers are distinct, the scaling regimes reveal themselves clearly. We observe that for large $Re$ the scaling exponent of $T_{RMS}/\Delta T$ with respect to $Re$ is essentially the same for both $Pr$. For $Re > 10^4$, temperature RMS exhibits the same scaling, $T_{RMS}/\Delta T \sim Re^{-0.2}$, for both $Pr$. However, the exponents in moderate Reynolds numbers, to the extent that they can be defined at all, are different, with $T_{RMS}/\Delta T$ showing $Re^{-0.13}$ and $Re^{-0.21}$ for $Pr = 0.1$ and $Pr = 1$, respectively. They are unlikely to be of fundamental significance.

\section{Fluctuation of global quantities}
\label{sec:balance}

\begin{figure}
\captionsetup{width=1\textwidth}
\centerline{
\includegraphics[width=1\textwidth]{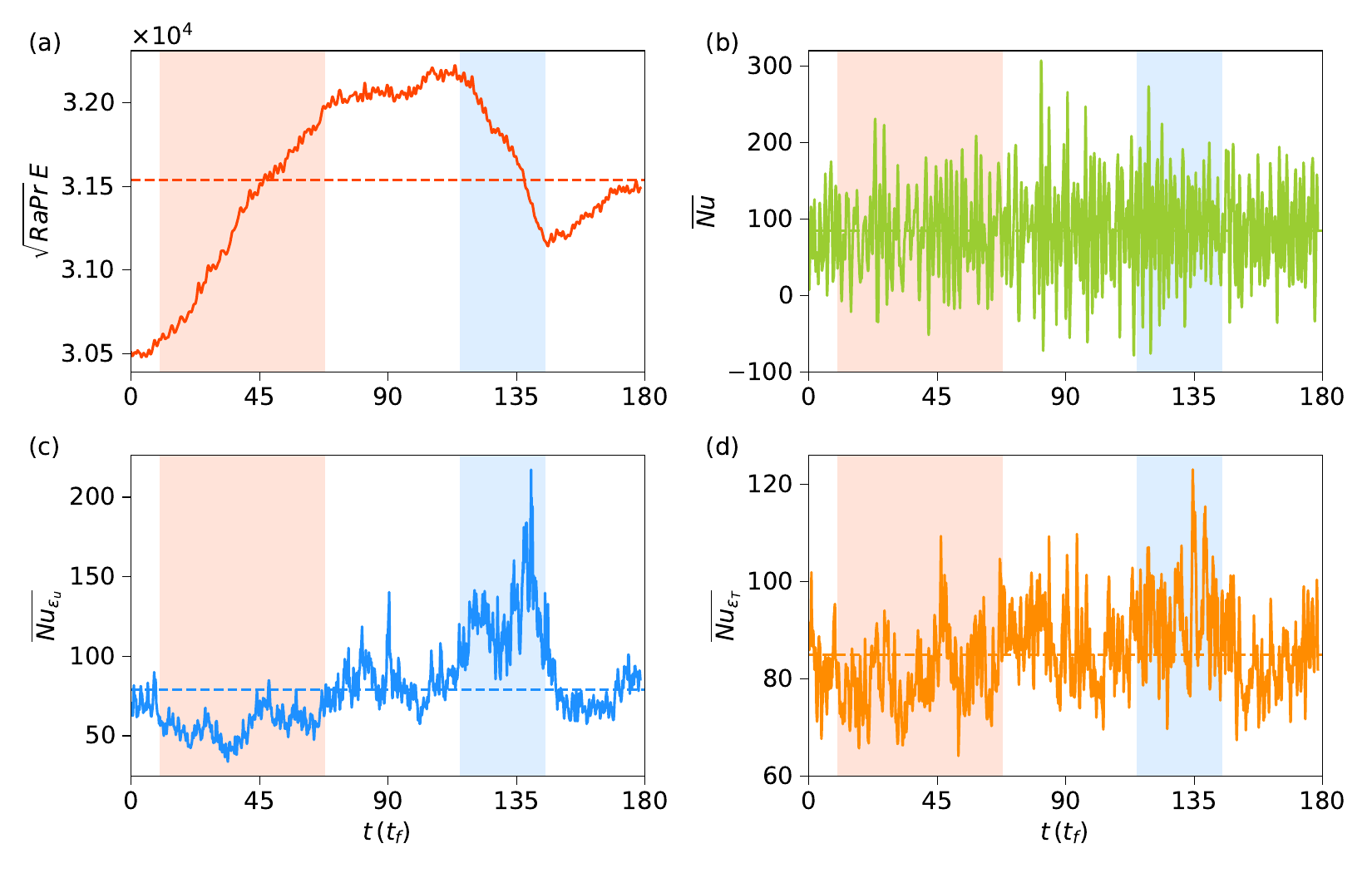}}
\caption{Temporal evolution of the integral quantities in statistically steady state for $Pr = 0.1, Ra = 10^{10}$. (a) Domain-averaged scaled kinetic energy $\sqrt{RaPr} \, E$ is dominated by a slow evolution. (b) $\overline{Nu}$ fluctuates rapidly about its mean. (c) $\overline{Nu_{\varepsilon_u}}$, in addition to having rapidly fluctuating components, evolves slowly and is related to $E$ [equation~\eqref{eq:E_bal_non}]. (d) $\overline{Nu_{\varepsilon_T}}$ fluctuates rapidly but a weak slowly-varying trend is present. The horizontal dashed line in all the panels indicates the time-averaged quantity. Here the origin is taken to be $3000 \, t_f$ of figure~\ref{fig:deter_trans}.}
\label{fig:fluct_time_Pr0.1}
\end{figure}

We now discuss the fluctuation of integral quantities in the nominally steady state. Taking the dot product of equation~\eqref{eq:u} with $\bm{u}$ and averaging over the entire domain, we obtain
\begin{equation}
\frac{\partial}{\partial t}  \langle u_i^2 / 2 \rangle_A  = \alpha g \langle u_z T \rangle_A - \langle  \varepsilon_u \rangle_A \, . \label{eq:E_bal_dim}
\end{equation}
Recalling that $E = \langle (u_x^2 + u_z^2)/2 \rangle_A/u_f^2$ is the domain-averaged kinetic energy, equation~\eqref{eq:E_bal_dim}
takes the non-dimensional form
\begin{equation}
\sqrt{RaPr} \frac{\partial E}{\partial t}  = \overline{Nu} - \overline{Nu_{\varepsilon_u}} \, , \label{eq:E_bal_non}
\end{equation}
where $\overline{Nu}$ and $\overline{Nu_{\varepsilon_u}}$ are the instantaneous domain-averaged heat fluxes defined in equations~\eqref{eq:Nu_t} and \eqref{eq:Nu_epsv_t}, respectively. Equation~\eqref{eq:E_bal_non} states that  $\overline{Nu}$ and $\overline{Nu_{\varepsilon_u}}$ are not equal to each other whenever $E$ varies with time.

In figure~\ref{fig:fluct_time_Pr0.1}, we show the temporal evolutions of $\sqrt{RaPr} E, \overline{Nu}, \overline{Nu_{\varepsilon_u}}$, and $\overline{Nu_{\varepsilon_T}}$ in the nominally steady state for $Pr = 0.1, Ra = 10^{10}$. Each quantity evolves differently from the other. Figure~\ref{fig:fluct_time_Pr0.1}(a) shows that domain-averaged kinetic energy $E$ contains sizable changes with respect to its mean value, indicated by a dashed horizontal line, occurring in 40-60 units of free-fall time. $\overline{Nu_{\varepsilon_u}}$ in figure~\ref{fig:fluct_time_Pr0.1}(c) shows a slow variation superimposed by strong rapid fluctuations. In contrast, $\overline{Nu}$ in figure~\ref{fig:fluct_time_Pr0.1}(b) fluctuates rapidly around its mean value. The $\overline{Nu_{\varepsilon_T}}$ in figure~\ref{fig:fluct_time_Pr0.1}(d) also fluctuates rapidly about its mean, with weak (but non-monotonic) trends. Data for $Ra > 10^8$ at $Pr = 0.1$ exhibit similar characteristics.

\begin{figure}
\captionsetup{width=1\textwidth}
\centerline{
\includegraphics[width=1\textwidth]{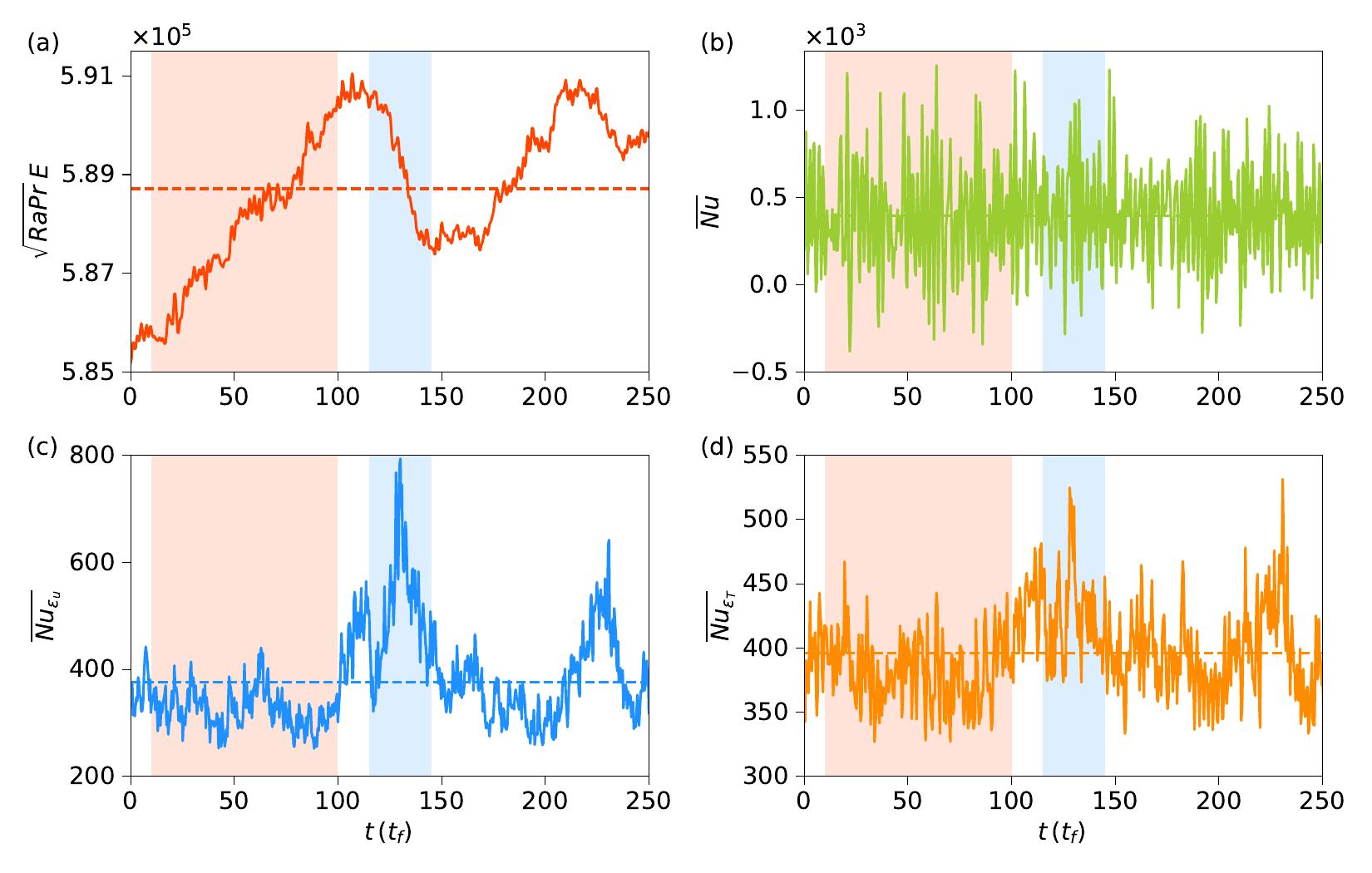}}
\caption{Temporal evolution of the integral quantities in steady state for $Pr = 1, Ra = 10^{12}$. The descriptions are the same as in figure~\ref{fig:fluct_time_Pr0.1}.}
\label{fig:fluct_time_Pr1}
\end{figure}

Coming to the energy balance equation, if we average equation~\eqref{eq:E_bal_non} over a finite interval of time, say between an initial (non-dimensional) time $t_{init}$ and a final time $t_{fin}$, we obtain
\begin{equation}
\sqrt{RaPr} \frac{\Delta E}{\Delta t} = \langle Nu \rangle_{\Delta t} - \langle Nu_{\varepsilon_u} \rangle_{\Delta t}, \label{eq:E_bal_fin}
\end{equation}
where $\Delta E = E(t_{fin}) - E(t_{init})$, $\Delta t = t_{fin} - t_{init}$, and $\langle \cdot \rangle_{\Delta t}$ denotes an averaging over the time interval $\Delta t$. As there exist long periods of growth or decay of $E$ (see figure~\ref{fig:fluct_time_Pr0.1}(a)), we can apply equation~\eqref{eq:E_bal_fin} to those intervals. For example, focusing on a segment where $E$ grows in figure~\ref{fig:fluct_time_Pr0.1} (the region highlighted by red shading), we find the LHS of equation~\eqref{eq:E_bal_fin} to be $\sqrt{RaPr} \Delta E/\Delta t = 23.8$. During this period the average values of the heat fluxes are found to be $\langle Nu \rangle_{\Delta t} = 80.3$, and $\langle Nu_{\varepsilon_u} \rangle_{\Delta t} = 56.4$, yielding 23.9 for the right hand side. Thus, equation~\eqref{eq:E_bal_fin} is satisfied perfectly. Similarly, in the blue-shaded region in figure~\ref{fig:fluct_time_Pr0.1} where $E$ decays, we obtain $\sqrt{RaPr} \Delta E/\Delta t = -33, \langle Nu \rangle_{\Delta t} = 90.4$, and $\langle Nu_{\varepsilon_u} \rangle_{\Delta t} = 123.5$; thus, the terms of equation~\eqref{eq:E_bal_fin} balance perfectly again. We note that the power spectra of the quantities shown in figure~\ref{fig:fluct_time_Pr0.1} show slow oscillations, or low-frequency modes, in $E$ and $Nu_{\varepsilon_u}$, $Nu$ and $Nu_{\varepsilon_T}$, although these modes diminish in power compared to the significantly strengthening high frequencies. 

On the one hand, due to the rapid fluctuation of $\overline{Nu}$ about its mean, its short-term average does not differ much from the long-term average. For example, $\langle Nu \rangle_{\Delta t} = 80.3$ and $\langle Nu \rangle_{\Delta t} = 90.4$ in the same two intervals are not far from the average of 84.7. On the other hand, the presence of a strong low frequency component in $\overline{Nu_{\varepsilon_u}}$ causes short-term averages to differ significantly from the long-term value. For example, $\langle Nu_{\varepsilon_u} \rangle_{\Delta t} = 56.4$ and $\langle Nu_{\varepsilon_u} \rangle_{\Delta t} = 123.5$ in the growing and decaying periods of $E$, respectively, differ by up to 50-60\% from $Nu_{\varepsilon_u} = 79.2$. This applies to all the high-$Re$ data in 2D RBC that we have explored. For example, figure~\ref{fig:fluct_time_Pr1} demonstrates the same picture for $Pr = 1, Ra = 10^{12}$ where, in the red- and blue-shaded regions, $\langle Nu \rangle_{\Delta t}$ is, respectively, larger and smaller than $\langle Nu_{\varepsilon_u} \rangle_{\Delta t}$, and equation~\eqref{eq:E_bal_fin} applies perfectly. 

As $Ra$ approaches very high values, the overwhelming resources required to simulate convective flows in two dimensions limit the total simulation time available to gather statistics. As a result, $Nu$ and $Nu_{\varepsilon_u}$ may not converge perfectly even if the sufficiency of spatial and temporal resolutions is ensured. For the simulation at $Pr = 0.1, Ra = 10^{10}$ (shown in figure~\ref{fig:fluct_time_Pr0.1}), $Nu$ and $Nu_{\varepsilon_u}$ differ by more than 6\% (see table~\ref{table:sim_detail}). Similarly, for $Pr = 1, Ra = 10^{12}$ (shown in figure~\ref{fig:fluct_time_Pr1}), the difference is also about 5\%. For lower $Ra$, on the other hand, there is better convergence to within 1-2\%. The convergence of $Nu_{\varepsilon_T}$ and $Nu$ is much better because both $\overline{Nu_{\varepsilon_T}}$ and $\overline{Nu}$ oscillate with comparable rapidity about their long-term averages.

\section{Summary and conclusions}

Our goal here has been to study the nature of the transient evolution of the DNS of 2D thermal convection, using the no-slip boundary condition on all the walls, along with isothermal bottom and top walls and adiabatic sidewalls. We illustrate related features using a square box for Prandtl numbers of 0.1 and 1, in the Rayleigh number range between $10^6$ and $10^{12}$. We particularly study the temporal evolution of integral transport quantities---such as the Nusselt number (defined in three different ways) and the turbulent energy---and discuss their scaling.  The nominally steady state is reached exponentially with substantial dependence on Rayleigh and Prandtl numbers. Although there is some degree of common behavior of transients for all the conditions explored here, there is no strict universality to the details of the exponential approach. We find, perhaps not surprisingly, that the velocity field evolves more slowly than the thermal field. That the velocity field evolves more slowly is not surprising because this evolution involves a few intermediate steps from the onset of heating changes. In the relation $Re \sim Ra^{\zeta}$, if the exponent $\zeta > 0.5$ as in 2D RBC, the ratio $u_{RMS}/u_f$ continues to increase with $Ra$. It is evident from equation~\eqref{eq:E} (see also \eqref{eq:E_growth} and \eqref{eq:t_trans}) that $t_{trns}$ increases with increasing fluctuating energy.

We also call attention to large oscillations of the velocity field in what may be regarded effectively as the steady state. One main conclusion is that these low-frequency oscillations are related to differences between the Nusselt number defined by the correlation of $u_z$ and $T$ and the Nusselt number based on the energy dissipation [see equation~\eqref{eq:E_bal_non}]. The time to saturation of the turbulent energy is presumably dependent on the formation of the large structure~\citep{Smith:PRL1993}, which itself would depend on the aspect ratio. The relation between the formation of the large structure and the time to saturation remains unclear at present, but it appears that achieving the so-called ultimate state of convection for smooth boundaries is as elusive in 2D as in 3D. The naive expectation that the reduced dimensionality in 2D allows one to settle the question of the ultimate state by attaining high enough $Ra$ is thwarted to some degree by the appearance of strong, low frequency fluctuations that demand excessively long simulation times for definitive scaling relations.


\backsection[Acknowledgements]{We appreciate long-term collaboration on convection studies with J\"org Schumacher. To him and to Erik Lindborg, Detlef Lohse, John Wettlaufer and Mahendra Verma, we are grateful for comments on an earlier draft. This research was carried out on the High Performance Computing resources at New York University Abu Dhabi. The authors also gratefully acknowledge {\sc Shaheen} II of KAUST, Saudi Arabia (under project nos. k1491 and k1624) for providing computational resources. }

\backsection[Funding]{This material is based upon work supported by Tamkeen under the NYU Abu Dhabi Research Institute grant G1502, and by the KAUST Office of Sponsored Research under Award URF/1/4342-01. AP also acknowledges financial support from SERB, India under the grant SRG/2023/001746 as well as from IIT Roorkee under FIG scheme. NYU supports KRS's research.}

\backsection[Declaration of Interests]{The authors report no conflict of interest.}

\backsection[Data availability statement]{The data that support the findings of this study are available from the corresponding author upon reasonable request.}

\backsection[Author ORCIDs]{\\
A. Pandey, https://orcid.org/0000-0001-8232-6626;\\
K. R. Sreenivasan, https://orcid.org/0000-0002-3943-6827}

\appendix
\section{Numerical details and effects of resolution on the transient state}
\label{sec:apndx}

As discussed in \S~\ref{sec:transient}, the transient time increases rapidly with $Ra$ in 2D RBC. Therefore, the steady state for high-$Ra$ flows is challenging to attain because the simulations require hundreds or thousands of free-fall times in the transient state, during which the domain-averaged kinetic energy continues to increase with time. Thus, in the transient flow state, conducting simulations with a mesh that resolves all relevant scales in the flow is extremely challenging due to a significant increase in the required computational resources and wait time. Therefore, we start the simulation with conduction temperature profile and random perturbations on a coarse mesh and continue until the domain-averaged kinetic energy stops growing with time and starts to fluctuate about some mean, whose value depends on $Ra$ and $Pr$. However, it is important to ensure that the steady state that is attained using a coarser mesh is close to the one that would be attained if a finer mesh is used. 

\begin{figure}
\captionsetup{width=1\textwidth}
\centerline{
\includegraphics[width=0.7\textwidth]{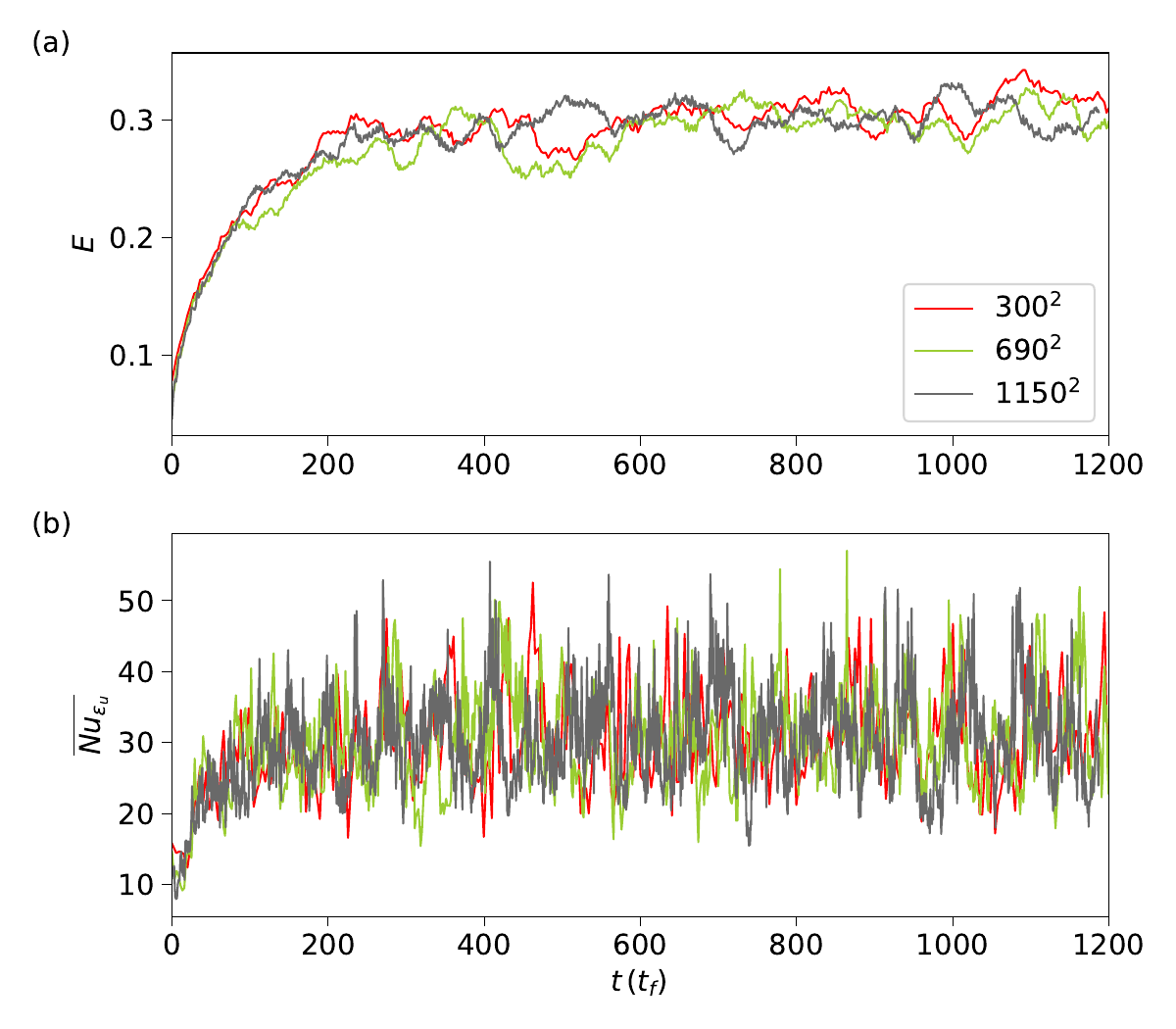}}
\caption{(a) Evolution of the domain-averaged energy $E$ for $Pr = 0.1, Ra = 3 \times 10^8$ using three different spatial resolutions. The similarity of the evolutions of $E$, in the statistical sense, suggests that the transient time and the mean energy in the steady state do not depend on the spatial resolution. (b) $\overline{Nu_{\varepsilon_u}}$ exhibits strong fluctuations, especially at moments when a rapid decay is observed in $E$. }
\label{fig:resolution_Pr0p1}
\end{figure}

To verify this, we performed simulations for a few governing parameters with different spatial resolutions and compared the temporal evolution of the integral quantities. In figure~\ref{fig:resolution_Pr0p1}, we show the evolution of $E$ for $Pr = 0.1, Ra = 3 \times 10^8$ in three different simulations, for mesh cells of $300^2$, $690^2$, and $1150^2$. We can see that the growth of $E$ in the initial stage (for $t < 120 \, t_f$) is similar in all simulations, although the evolutions differ slightly for intermediate stages. However, once $E$ stops growing and starts fluctuating about some mean, the average value of energy in the steady state does not depend on the spatial resolution. We find that the mean energy for $t > 400 t_f$ in figure~\ref{fig:resolution_Pr0p1} differs only by at most 2\%. We also show the evolution of $\overline{Nu_{\varepsilon_u}}$ in figure~\ref{fig:resolution_Pr0p1}(b) for the three simulations and observe wild fluctuations, especially when $E$ decreases rapidly over a short time interval. Thus, the flow properties in the steady state seem to be largely unaffected by the spatial resolution used in the transient state. 

\begin{figure}
\captionsetup{width=1\textwidth}
\centerline{
\includegraphics[width=0.6\textwidth]{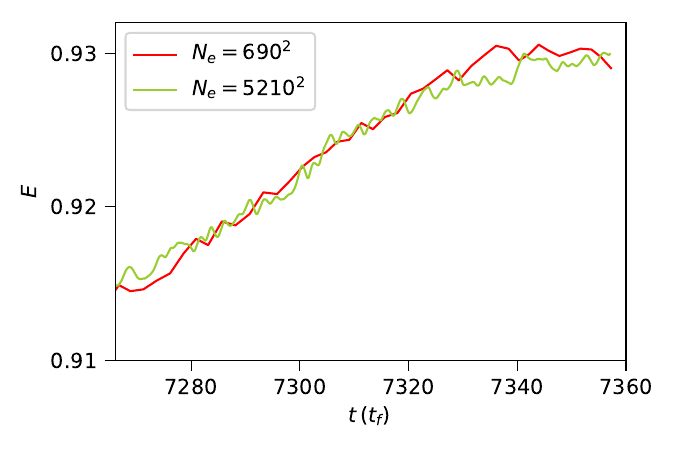}}
\caption{Evolution of the kinetic energy in the intermediate stage of the transient state for $Pr = 1, Ra = 10^{12}$. Computational time is close to one million core-hours for the simulation with $5210^2$ mesh cells---much bigger than one thousand core hours needed for simulation with $690^2$ cells. It is clear that performing high-$Ra$ simulation in the transient state with full resolution is infeasible.}
\label{fig:resolution_Pr1}
\end{figure}

Similarly, figure~\ref{fig:resolution_Pr1} shows $E$ for intermediate stages of the transient state for $Pr = 1, Ra = 10^{12}$. We start the simulation with a coarse resolution having $690^2$ mesh cells (red curve). While $E$ is still growing, we ramp up the resolution and start another simulation with $5210^2$ mesh cells (green curve), in addition to continuing the original one. Figure~\ref{fig:resolution_Pr1} shows that the trajectories of $E$ in both simulations are similar, while the computational resources differ substantially. On the one hand, simulation with the coarse mesh consumes only one thousand core hours, and the segment shown in figure~\ref{fig:resolution_Pr1} was obtained in just eight hours on 128 cores. On the other hand, simulation with the finer mesh takes more than 900 hours on 1024 cores (green curve). As the transient time is nearly $8000 \, t_f$ for these parameters (see figure~\ref{fig:t_trns}), one would need to run the simulation with $5210^2$ mesh cells for 72000 hours on 1024 cores to attain the steady state, which is clearly impossible at present.


\begin{thebibliography}{54}
\expandafter\ifx\csname natexlab\endcsname\relax\def\natexlab#1{#1}\fi
\def\au#1{#1} \def\ed#1{#1} \def\yr#1{#1}\def\at#1{#1}\def\jt#1{\textit{#1}}
  \def\bt#1{#1}\def\bvol#1{\textbf{#1}} \def\vol#1{#1} \def\pg#1{#1}
  \def\publ#1{#1}\def\arxiv#1{#1}\def\org#1{#1}\def\st#1{\textit{#1}}

\bibitem[{Castaing} {\em et~al.\/}(1989){Castaing}, {Gunaratne}, {Kadanoff},
  {Libchaber} \& {Heslot}]{Castaing:JFM1989}
{\sc \au{{Castaing}, B.}, \au{{Gunaratne}, G.}, \au{{Kadanoff}, L.},
  \au{{Libchaber}, A.} \& \au{{Heslot}, F.}} \yr{1989}  \at{Scaling of hard
  thermal turbulence in {R}ayleigh-{B}{\'e}nard convection}.  \jt{J. Fluid
  Mech.}  \bvol{204},  \pg{1--30}.

\bibitem[{Chill\`{a}} \& {Schumacher}(2012)]{Chilla:EPJE2012}
{\sc \au{{Chill\`{a}}, F.} \& \au{{Schumacher}, J.}} \yr{2012}  \at{New
  perspectives in turbulent {R}ayleigh-{B}{\'e}nard convection}.  \jt{Eur.
  Phys. J. E}  \bvol{35},  \pg{58}.

\bibitem[Chini \& Cox(2009)]{Chini:PoF2009}
{\sc \au{Chini, G.~P.} \& \au{Cox, S.~M.}} \yr{2009}  \at{Large {R}ayleigh
  number thermal convection: Heat flux predictions and strongly nonlinear
  solutions}.  \jt{Phys. Fluids}  \bvol{21}~(8),  \pg{083603}.

\bibitem[Doering(2020)]{Doering:PRL2020}
{\sc \au{Doering, C.~R.}} \yr{2020}  \at{Absence of evidence for the ultimate
  state of turbulent {R}ayleigh-{B}\'enard convection}.  \jt{Phys. Rev. Lett.}
  \bvol{124},  \pg{229401}.

\bibitem[Doering {\em et~al.\/}(2019)Doering, Toppaladoddi \&
  Wettlaufer]{Doering:PRL2019}
{\sc \au{Doering, C.~R.}, \au{Toppaladoddi, S.} \& \au{Wettlaufer, J.~S.}}
  \yr{2019}  \at{Absence of evidence for the ultimate regime in two-dimensional
  {R}ayleigh-{B}\'enard convection}.  \jt{Phys. Rev. Lett.}  \bvol{123},
  \pg{259401}.

\bibitem[Foroozani {\em et~al.\/}(2014)Foroozani, Niemela, Armenio \&
  Sreenivasan]{Foroozani:PRE2014}
{\sc \au{Foroozani, N.}, \au{Niemela, J.~J.}, \au{Armenio, V.} \&
  \au{Sreenivasan, K.~R.}} \yr{2014}  \at{Influence of container shape on
  scaling of turbulent fluctuations in convection}.  \jt{Phys. Rev. E}
  \bvol{90},  \pg{063003}.

\bibitem[Foroozani {\em et~al.\/}(2017)Foroozani, Niemela, Armenio \&
  Sreenivasan]{Foroozani:PRE2017}
{\sc \au{Foroozani, N.}, \au{Niemela, J.~J.}, \au{Armenio, V.} \&
  \au{Sreenivasan, K.~R.}} \yr{2017}  \at{Reorientations of the large-scale
  flow in turbulent convection in a cube}.  \jt{Phys. Rev. E}  \bvol{95},
  \pg{033107}.

\bibitem[Gao {\em et~al.\/}(2024)Gao, Tao, Huang, Bao \& Xie]{Gao:JFM2024}
{\sc \au{Gao, Z.-Y.}, \au{Tao, X.}, \au{Huang, S.-D.}, \au{Bao, Y.} \& \au{Xie,
  Y.-C.}} \yr{2024}  \at{Flow state transition induced by emergence of orbiting
  satellite eddies in two-dimensional turbulent {R}ayleigh-{B}\'enard
  convection}.  \jt{J. Fluid Mech.}  \bvol{997},  \pg{A54}.

\bibitem[{Grossmann} \& {Lohse}(2000)]{Grossmann:JFM2000}
{\sc \au{{Grossmann}, S.} \& \au{{Lohse}, D.}} \yr{2000}  \at{Scaling in
  thermal convection: a unifying theory}.  \jt{J. Fluid Mech.}  \bvol{407},
  \pg{27–56}.

\bibitem[Grossmann \& Lohse(2001)]{Grossmann:PRL2001}
{\sc \au{Grossmann, S.} \& \au{Lohse, D.}} \yr{2001}  \at{Thermal convection
  for large {P}randtl numbers}.  \jt{Phys. Rev. Lett.}  \bvol{86},
  \pg{3316--3319}.

\bibitem[Howard(1972)]{Howard:ARFM1972}
{\sc \au{Howard, L.~N.}} \yr{1972}  \at{Bounds on flow quantities}.  \jt{Annu.
  Rev. Fluid Mech.}  \bvol{4}~(1),  \pg{473--494},  \arxiv{arXiv:
  https://doi.org/10.1146/annurev.fl.04.010172.002353}.

\bibitem[Iyer {\em et~al.\/}(2020)Iyer, Scheel, Schumacher \&
  Sreenivasan]{Iyer:PNAS2020}
{\sc \au{Iyer, K.~P.}, \au{Scheel, J.~D.}, \au{Schumacher, J.} \&
  \au{Sreenivasan, K.~R.}} \yr{2020}  \at{Classical 1/3 scaling of convection
  holds up to {R}a = $10^{15}$}.  \jt{Proc. Natl. Acad. Sci. USA}
  \bvol{117}~(14),  \pg{7594--7598},  \arxiv{arXiv:
  https://www.pnas.org/content/117/14/7594.full.pdf}.

\bibitem[{Johnston} \& {Doering}(2009)]{Johnston:PRL2009}
{\sc \au{{Johnston}, H.} \& \au{{Doering}, C.~R.}} \yr{2009}  \at{Comparison of
  turbulent thermal convection between conditions of constant temperature and
  constant flux}.  \jt{Phys. Rev. Lett.}  \bvol{102},  \pg{064501}.

\bibitem[Kadanoff(2001)]{Kadanoff:PT2001}
{\sc \au{Kadanoff, L.~P.}} \yr{2001}  \at{Turbulent heat flow: Structures and
  scaling}.  \jt{Phys. Today}  \bvol{54}~(8),  \pg{34--39},  \arxiv{arXiv:
  https://pubs.aip.org/physicstoday/article-pdf/54/8/34/16746047/34\_1\_online.pdf}.

\bibitem[Labarre {\em et~al.\/}(2023)Labarre, Fauve \&
  Chibbaro]{Labarre:PRF2023}
{\sc \au{Labarre, V.}, \au{Fauve, S.} \& \au{Chibbaro, S.}} \yr{2023}
  \at{Heat-flux fluctuations revealing regime transitions in
  {R}ayleigh-{B}\'enard convection}.  \jt{Phys. Rev. Fluids}  \bvol{8},
  \pg{053501}.

\bibitem[{Lam} {\em et~al.\/}(2002){Lam}, {Shang}, {Zhou} \&
  {Xia}]{Lam:PRE2002}
{\sc \au{{Lam}, S.}, \au{{Shang}, X.-D.}, \au{{Zhou}, S.-Q.} \& \au{{Xia},
  K.-Q.}} \yr{2002}  \at{{P}randtl number dependence of the viscous boundary
  layer and the {R}eynolds numbers in {R}ayleigh-{B}{\'e}nard convection}.
  \jt{Phys. Rev. E}  \bvol{65},  \pg{066306}.

\bibitem[Lohse \& Shishkina(2023)]{Lohse:PT2023}
{\sc \au{Lohse, D.} \& \au{Shishkina, O.}} \yr{2023}  \at{Ultimate turbulent
  thermal convection}.  \jt{Phys. Today}  \bvol{76}~(11),  \pg{26--32},
  \arxiv{arXiv:
  https://pubs.aip.org/physicstoday/article-pdf/76/11/26/20085578/26\_1\_pt.3.5341.pdf}.

\bibitem[Lohse \& Shishkina(2024)]{Lohse:RMP2024}
{\sc \au{Lohse, D.} \& \au{Shishkina, O.}} \yr{2024}  \at{Ultimate
  {R}ayleigh-{B}\'enard turbulence}.  \jt{Rev. Mod. Phys.}  \bvol{96},
  \pg{035001}.

\bibitem[{Niemela} {\em et~al.\/}(2000){Niemela}, {Skrbek}, {Sreenivasan} \&
  {Donnelly}]{Niemela:Nature2000}
{\sc \au{{Niemela}, J.~J.}, \au{{Skrbek}, L.}, \au{{Sreenivasan}, K.~R.} \&
  \au{{Donnelly}, R.~J.}} \yr{2000}  \at{Turbulent convection at very high
  {R}ayleigh numbers}.  \jt{Nature}  \bvol{404},  \pg{837--840}.

\bibitem[Pandey(2021)]{Pandey:JFM2021}
{\sc \au{Pandey, A.}} \yr{2021}  \at{Thermal boundary layer structure in
  low-{P}randtl-number turbulent convection}.  \jt{J. Fluid Mech.}  \bvol{910},
   \pg{A13}.

\bibitem[Pandey {\em et~al.\/}(2022{\natexlab{{\em a\/}}})Pandey, Krasnov,
  Schumacher, Samtaney \& Sreenivasan]{Pandey:PD2022}
{\sc \au{Pandey, A.}, \au{Krasnov, D.}, \au{Schumacher, J.}, \au{Samtaney, R.}
  \& \au{Sreenivasan, K.~R.}} \yr{2022{\natexlab{{\em a\/}}}}  \at{Similarities
  between characteristics of convective turbulence in confined and extended
  domains}.  \jt{Physica D}  \bvol{442},  \pg{133537}.

\bibitem[Pandey {\em et~al.\/}(2022{\natexlab{{\em b\/}}})Pandey, Krasnov,
  Sreenivasan \& Schumacher]{Pandey:JFM2022}
{\sc \au{Pandey, A.}, \au{Krasnov, D.}, \au{Sreenivasan, K.~R.} \&
  \au{Schumacher, J.}} \yr{2022{\natexlab{{\em b\/}}}}  \at{Convective
  mesoscale turbulence at very low {P}randtl numbers}.  \jt{J. Fluid Mech.}
  \bvol{948},  \pg{A23}.

\bibitem[Pandey {\em et~al.\/}(2016)Pandey, Kumar, Chatterjee \&
  Verma]{Pandey:PRE2016}
{\sc \au{Pandey, A.}, \au{Kumar, A.}, \au{Chatterjee, A.~G.} \& \au{Verma,
  M.~K.}} \yr{2016}  \at{Dynamics of large-scale quantities in
  {R}ayleigh-{B}\'enard convection}.  \jt{Phys. Rev. E}  \bvol{94},
  \pg{053106}.

\bibitem[Pandey \& Sreenivasan(2021)]{Pandey:EPL2021}
{\sc \au{Pandey, A.} \& \au{Sreenivasan, K.~R.}} \yr{2021}  \at{Convective heat
  transport in slender cells is close to that in wider cells at high {R}ayleigh
  and {P}randtl numbers}.  \jt{Europhys. Lett.}  \bvol{135}~(2),  \pg{24001}.

\bibitem[Pandey \& Verma(2016)]{Pandey:PoF2016}
{\sc \au{Pandey, A.} \& \au{Verma, M.~K.}} \yr{2016}  \at{Scaling of
  large-scale quantities in {R}ayleigh-{B}{\'e}nard convection}.  \jt{Phys.
  Fluids}  \bvol{28}~(9),  \pg{095105},  \arxiv{arXiv:
  https://doi.org/10.1063/1.4962307}.

\bibitem[{Pandey} {\em et~al.\/}(2016){Pandey}, {Verma}, {Chatterjee} \&
  {Dutta}]{Pandey:Pramana2016}
{\sc \au{{Pandey}, A.}, \au{{Verma}, M.~K.}, \au{{Chatterjee}, A.~G.} \&
  \au{{Dutta}, B.}} \yr{2016}  \at{Similarities between 2{D} and 3{D}
  convection for large {P}randtl number}.  \jt{Pramana - J. Phys.}  \bvol{87},
  \pg{13}.

\bibitem[Samuel \& Verma(2024)]{Samuel:PRF2024}
{\sc \au{Samuel, R.} \& \au{Verma, M.~K.}} \yr{2024}  \at{{B}olgiano-{O}bukhov
  scaling in two-dimensional {R}ayleigh-{B}\'enard convection at extreme
  {R}ayleigh numbers}.  \jt{Phys. Rev. Fluids}  \bvol{9},  \pg{023502}.

\bibitem[Samuel {\em et~al.\/}(2024)Samuel, Bode, Scheel, Sreenivasan \&
  Schumacher]{Samuel:JFM2024}
{\sc \au{Samuel, R.~J.}, \au{Bode, M.}, \au{Scheel, J.~D.}, \au{Sreenivasan,
  K.~R.} \& \au{Schumacher, J.}} \yr{2024}  \at{No sustained mean velocity in
  the boundary region of plane thermal convection}.  \jt{J. Fluid Mech.}
  \bvol{996},  \pg{A49}.

\bibitem[{Scheel} {\em et~al.\/}(2013){Scheel}, {Emran} \&
  {Schumacher}]{Scheel:NJP2013}
{\sc \au{{Scheel}, J.~D.}, \au{{Emran}, M.~S.} \& \au{{Schumacher}, J.}}
  \yr{2013}  \at{Resolving the fine-scale structure in turbulent
  {R}ayleigh-{B}{\'e}nard convection}.  \jt{New J. Phys.}  \bvol{15},
  \pg{113063}.

\bibitem[Scheel {\em et~al.\/}(2012)Scheel, Kim \& White]{Scheel:JFM2012}
{\sc \au{Scheel, J.~D.}, \au{Kim, E.} \& \au{White, K.~R.}} \yr{2012}
  \at{Thermal and viscous boundary layers in turbulent {R}ayleigh–{B}\'enard
  convection}.  \jt{J. Fluid Mech.}  \bvol{711},  \pg{281–305}.

\bibitem[Scheel \& Schumacher(2016)]{Scheel:JFM2016}
{\sc \au{Scheel, J.~D.} \& \au{Schumacher, J.}} \yr{2016}  \at{Global and local
  statistics in turbulent convection at low {P}randtl numbers}.  \jt{J. Fluid
  Mech.}  \bvol{802},  \pg{147–173}.

\bibitem[{Schmalzl} {\em et~al.\/}(2004){Schmalzl}, {Breuer} \&
  {Hansen}]{Schmalzl:EPL2004}
{\sc \au{{Schmalzl}, J.}, \au{{Breuer}, M.} \& \au{{Hansen}, U.}} \yr{2004}
  \at{On the validity of two-dimensional numerical approaches to time-dependent
  thermal convection}.  \jt{Europhys. Lett.}  \bvol{67},  \pg{390--396}.

\bibitem[{Schumacher} {\em et~al.\/}(2015){Schumacher}, {G\"{o}tzfried} \&
  {Scheel}]{Schumacher:PNAS2015}
{\sc \au{{Schumacher}, J.}, \au{{G\"{o}tzfried}, P.} \& \au{{Scheel}, J.~D.}}
  \yr{2015}  \at{Enhanced enstrophy generation for turbulent convection in
  low-{P}randtl-number fluids}.  \jt{Proc. Natl. Acad. Sci. USA}  \bvol{112},
  \pg{9530--9535}.

\bibitem[Schumacher \& Sreenivasan(2020)]{Schumacher:RMP2020}
{\sc \au{Schumacher, J.} \& \au{Sreenivasan, K.~R.}} \yr{2020}  \at{Colloquium:
  Unusual dynamics of convection in the {S}un}.  \jt{Rev. Mod. Phys.}
  \bvol{92},  \pg{041001}.

\bibitem[Shraiman \& Siggia(1990)]{Shraiman:PRA1990}
{\sc \au{Shraiman, B.~I.} \& \au{Siggia, E.~D.}} \yr{1990}  \at{Heat transport
  in high-{R}ayleigh-number convection}.  \jt{Phys. Rev. A}  \bvol{42},
  \pg{3650--3653}.

\bibitem[Siggia(1994)]{Siggia:ARFM1994}
{\sc \au{Siggia, E.~D.}} \yr{1994}  \at{High {R}ayleigh number convection}.
  \jt{Annu. Rev. Fluid Mech.}  \bvol{26}~(1),  \pg{137--168},  \arxiv{arXiv:
  https://doi.org/10.1146/annurev.fl.26.010194.001033}.

\bibitem[Smith \& Yakhot(1993)]{Smith:PRL1993}
{\sc \au{Smith, L.~M.} \& \au{Yakhot, V.}} \yr{1993}  \at{Bose condensation and
  small-scale structure generation in a random force driven 2{D} turbulence}.
  \jt{Phys. Rev. Lett.}  \bvol{71},  \pg{352--355}.

\bibitem[Sreenivasan {\em et~al.\/}(2002)Sreenivasan, Bershadskii \&
  Niemela]{Sreenivasan:PRE2002}
{\sc \au{Sreenivasan, K.~R.}, \au{Bershadskii, A.} \& \au{Niemela, J.~J.}}
  \yr{2002}  \at{Mean wind and its reversal in thermal convection}.  \jt{Phys.
  Rev. E}  \bvol{65},  \pg{056306}.

\bibitem[{Stevens} {\em et~al.\/}(2011){Stevens}, {Lohse} \&
  {Verzicco}]{Stevens:JFM2011}
{\sc \au{{Stevens}, R.}, \au{{Lohse}, D.} \& \au{{Verzicco}, R.}} \yr{2011}
  \at{{P}randtl and {R}ayleigh number dependence of heat transport in high
  {R}ayleigh number thermal convection}.  \jt{J. Fluid Mech.}  \bvol{688},
  \pg{31--43}.

\bibitem[{Stevens} {\em et~al.\/}(2010){Stevens}, {Verzicco} \&
  {Lohse}]{Stevens:JFM2010}
{\sc \au{{Stevens}, R.}, \au{{Verzicco}, R.} \& \au{{Lohse}, D.}} \yr{2010}
  \at{Radial boundary layer structure and {N}usselt number in
  {R}ayleigh-{B}{\'e}nard convection}.  \jt{J. Fluid Mech.}  \bvol{643},
  \pg{495--507}.

\bibitem[Stevens {\em et~al.\/}(2024)Stevens, Hartmann, Verzicco \&
  Lohse]{Stevens:JFM2024}
{\sc \au{Stevens, R.~J.}, \au{Hartmann, R.}, \au{Verzicco, R.} \& \au{Lohse,
  D.}} \yr{2024}  \at{How wide must {R}ayleigh–{B}\'enard cells be to prevent
  finite aspect ratio effects in turbulent flow?}  \jt{J. Fluid Mech.}
  \bvol{1000},  \pg{A58}.

\bibitem[Tiwari {\em et~al.\/}(2025)Tiwari, Sharma \& Verma]{Tiwari:IJHMT2025}
{\sc \au{Tiwari, H.}, \au{Sharma, L.} \& \au{Verma, M.~K.}} \yr{2025}
  \at{Compressible turbulent convection at very high {R}ayleigh numbers}.
  \jt{Int. J. Heat Mass Transfer}  \bvol{242},  \pg{126821}.

\bibitem[{van der Poel} {\em et~al.\/}(2011){van der Poel}, {Stevens} \&
  {Lohse}]{Poel:PRE2011}
{\sc \au{{van der Poel}, E.~P.}, \au{{Stevens}, R. J. A.~M.} \& \au{{Lohse},
  D.}} \yr{2011}  \at{Connecting flow structures and heat flux in turbulent
  {R}ayleigh-{B}{\'e}nard convection}.  \jt{Phys. Rev. E}  \bvol{84},
  \pg{045303(R)}.

\bibitem[{van der Poel} {\em et~al.\/}(2013){van der Poel}, {Stevens} \&
  {Lohse}]{Poel:JFM2013}
{\sc \au{{van der Poel}, E.~P.}, \au{{Stevens}, R. J. A.~M.} \& \au{{Lohse},
  D.}} \yr{2013}  \at{Comparison between two- and three-dimensional
  {R}ayleigh-{B}{\'e}nard convection}.  \jt{J. Fluid Mech.}  \bvol{736},
  \pg{177--194}.

\bibitem[Verma(2018)]{Verma:book2018}
{\sc \au{Verma, M.~K.}} \yr{2018} {\em Physics of Buoyant Flows\/}.
  \publ{Singapore: World Scientific},  \arxiv{arXiv:
  https://www.worldscientific.com/doi/pdf/10.1142/10928}.

\bibitem[Verma {\em et~al.\/}(2017)Verma, Kumar \& Pandey]{Verma:NJP2017}
{\sc \au{Verma, M.~K.}, \au{Kumar, A.} \& \au{Pandey, A.}} \yr{2017}
  \at{Phenomenology of buoyancy-driven turbulence: recent results}.  \jt{New J.
  Phys.}  \bvol{19}~(2),  \pg{025012}.

\bibitem[Verzicco \& Camussi(1999)]{Verzicco:JFM1999}
{\sc \au{Verzicco, R.} \& \au{Camussi, R.}} \yr{1999}  \at{{P}randtl number
  effects in convective turbulence}.  \jt{J. Fluid Mech.}  \bvol{383},
  \pg{55–73}.

\bibitem[Wan {\em et~al.\/}(2020)Wan, Wang, Wang, Xia, Zhou \&
  Sun]{Wan:JFM2020}
{\sc \au{Wan, Z.-H.}, \au{Wang, Q.}, \au{Wang, B.}, \au{Xia, S.-N.}, \au{Zhou,
  Q.} \& \au{Sun, D.-J.}} \yr{2020}  \at{On non-{O}berbeck–{B}oussinesq
  effects in {R}ayleigh-{B}\'enard convection of air for large temperature
  differences}.  \jt{J. Fluid Mech.}  \bvol{889},  \pg{A10}.

\bibitem[Weiss \& Ahlers(2011)]{Weiss:JFM2011}
{\sc \au{Weiss, S.} \& \au{Ahlers, G.}} \yr{2011}  \at{Turbulent
  {R}ayleigh–{B}\'enard convection in a cylindrical container with aspect
  ratio $\gamma$ = 0.50 and {P}randtl number {P}r = 4.38}.  \jt{J. Fluid Mech.}
   \bvol{676},  \pg{5–40}.

\bibitem[Wen {\em et~al.\/}(2020)Wen, Goluskin, LeDuc, Chini \&
  Doering]{Wen:JFM2020}
{\sc \au{Wen, B.}, \au{Goluskin, D.}, \au{LeDuc, M.}, \au{Chini, G.~P.} \&
  \au{Doering, C.~R.}} \yr{2020}  \at{Steady {R}ayleigh-{B}\'enard convection
  between stress-free boundaries}.  \jt{J. Fluid Mech.}  \bvol{905},  \pg{R4}.

\bibitem[{Xi} \& {Xia}(2008)]{Xi:POF2008}
{\sc \au{{Xi}, H.} \& \au{{Xia}, K.}} \yr{2008}  \at{Flow mode transitions in
  turbulent thermal convection}.  \jt{Phys. Fluids}  \bvol{20},  \pg{5104}.

\bibitem[{Xia} {\em et~al.\/}(2003){Xia}, {Sun} \& {Zhou}]{Xia:PRE2003}
{\sc \au{{Xia}, K.~Q.}, \au{{Sun}, C.} \& \au{{Zhou}, S.~Q.}} \yr{2003}
  \at{Particle image velocimetry measurement of the velocity field in turbulent
  thermal convection}.  \jt{Phys. Rev. E}  \bvol{68},  \pg{066303}.

\bibitem[Zhang {\em et~al.\/}(2017)Zhang, Zhou \& Sun]{Zhang:JFM2017}
{\sc \au{Zhang, Y.}, \au{Zhou, Q.} \& \au{Sun, C.}} \yr{2017}  \at{Statistics
  of kinetic and thermal energy dissipation rates in two-dimensional turbulent
  {R}ayleigh-{B}{\'e}nard convection}.  \jt{J. Fluid Mech.}  \bvol{814},
  \pg{165–184}.

\bibitem[Zhu {\em et~al.\/}(2018)Zhu, Mathai, Stevens, Verzicco \&
  Lohse]{Zhu:PRL2018}
{\sc \au{Zhu, X.}, \au{Mathai, V.}, \au{Stevens, R. J. A.~M.}, \au{Verzicco,
  R.} \& \au{Lohse, D.}} \yr{2018}  \at{Transition to the ultimate regime in
  two-dimensional {R}ayleigh-{B}\'enard convection}.  \jt{Phys. Rev. Lett.}
  \bvol{120},  \pg{144502}.

\end{thebibliography}
\end{document}